\documentclass{lmcs}
\pdfoutput=1

\usepackage{lastpage}
\lmcsdoi{18}{1}{26}
\lmcsheading{}{\pageref{LastPage}}{}{}%
{Dec.~20,~2020}{Feb.~03,~2022}{}


\usepackage[utf8]{inputenc}
\usepackage[T1]{fontenc}

\usepackage{amsmath, amsthm, amssymb}

\usepackage[absolute]{textpos}

\usepackage{comment}

\usepackage{thmtools}
\usepackage{thm-restate}

\usepackage{algpseudocode}

\usepackage{xspace}

\usepackage{xfrac}
\usepackage{stmaryrd}

\def\cqedsymbol{\ifmmode$\lrcorner$\else{\unskip\nobreak\hfil
\penalty50\hskip1em\null\nobreak\hfil$\lrcorner$
\parfillskip=0pt\finalhyphendemerits=0\endgraf}\fi} 

\newcommand{\cqed}{\renewcommand{\qed}{\cqedsymbol}}

\newcommand{\executeiffilenewer}[3]{%
\ifnum\pdfstrcmp{\pdffilemoddate{#1}}%
{\pdffilemoddate{#2}}>0%
{\immediate\write18{#3}}\fi%
} 
\newcommand{\Oh}{\ensuremath{\mathcal{O}}}

\newcommand{\tw}{\ensuremath{\mathtt{tw}}\xspace}

\newcommand{\pmax}{\mathsf{pmax}}
\newcommand{\pmin}{\mathsf{pmin}}
\newcommand{\wsum}{\mathsf{sum}}

\newcommand{\mso}{{\sc mso}\xspace}

\newcommand{\set}[1]{\{#1\}}
\newcommand{\Nat}{\mathbb N}

\newcommand{\bag}{\mathsf{bag}}
\newcommand{\adh}{\mathsf{adh}}
\newcommand{\comp}{\mathsf{cmp}}
\newcommand{\mar}{\mathsf{mrg}}

\newcommand{\refine}{\mathsf{fact}}

\newcommand{\Jj}{\mathcal{J}}
\newcommand{\Ii}{\mathcal{I}}
\newcommand{\As}{\mathbb{A}}

\title{Optimizing tree decompositions in MSO}

\author[M.~Boja\'nczyk]{Miko\l{}aj Boja\'nczyk}

\author[M.~Pilipczuk]{Micha\l{} Pilipczuk}
\address{Institute of Informatics, University of Warsaw, Poland}
\email{bojan@mimuw.edu.pl, michal.pilipczuk@mimuw.edu.pl}

\thanks{An extended abstract of this work appeared in the proceedings of the $34^{\textrm{th}}$ Symposium on Theoretical Aspects of Computer Science, STACS
2017~\cite{BojanczykP17}. The research of M.~Bojańczyk is supported by the European
    Research Council (ERC) under the European Union's Horizon 2020 research
    and innovation programme (ERC consolidator grant LIPA, agreement no.\ 683080). 
During the work on this project, Mi.~Pilipczuk was supported by the Foundation for Polish Science via the START stipend programme.}

\begin{document}
 
\maketitle

\begin{abstract}
The classic algorithm of Bodlaender and Kloks~[J. Algorithms, 1996] solves the following problem in linear fixed-parameter time: 
given a tree decomposition of a graph of (possibly suboptimal) width~$k$, compute an optimum-width tree decomposition of the graph.
In this work, we prove that this problem can also be solved in \mso in the following sense: 
for every positive integer $k$, there is an \mso transduction from tree decompositions of width~$k$ to tree decompositions of optimum width.
Together with our recent results~[LICS~2016], this implies that for every~$k$ there exists an \mso transduction
which inputs a graph of treewidth~$k$, and nondeterministically outputs its tree decomposition of optimum width.
We also show that \mso transductions can be implemented in linear fixed-parameter time, which enables us to derive
the algorithmic result of Bodlaender and Kloks as a corollary of our main result.

\end{abstract}

\section{Introduction}

Consider the following problem: given a tree decomposition of a graph of some width $k$, possibly suboptimal, we would like to compute an optimum-width tree decomposition of the graph.
A classic algorithm of Bodlaender and Kloks~\cite{BodlaenderK96} solves this problem in linear fixed-parameter time complexity, where the input width $k$ is the parameter.

\begin{thm}[Bodlaender and Kloks,~\cite{BodlaenderK96}]\label{thm:bodlaender-kloks}
There exists an algorithm that, 
given a graph~$G$ on $n$ vertices and its tree decomposition of width $k$,
runs in time $2^{\Oh(k^3)}\cdot n$ and returns a tree decomposition of $G$ of optimum width.
\end{thm}

The algorithm of Bodlaender and Kloks applies a dynamic programming procedure that processes the input decomposition in a bottom-up manner. For every subtree, a set of partial optimum-width decompositions
is computed. The crucial ingredient is a combinatorial analysis of partial decompositions which shows that only some small subset of them, of size bounded only by a function of $k$, needs to be remembered for future computation.

The central component of this analysis is the notion of {\em{typical sequences}}: a specific normal form of integer sequences that is used to bound the complexity of partial decompositions useful for extensions. This notion was independently introduced by Bodlander and Kloks~\cite{BodlaenderK91} and by Lagergren and Arnborg~\cite{LagergrenA91} (the latter called them {\em{non-redundant sequences}}). Typical sequences have found applications in algorithms for computing other width measures,
like branchwidth~\cite{BodlaenderT97}, cutwidth~\cite{ThilikosSB05,ThilikosSB05a}, as well as pathwidth and branchwidth of matroids~\cite{Jeong0O16,Jeong0O18}.
Also, they were used in the context of finite-state recognizability of properties of graphs of bounded treewidth~\cite{CourcelleL96}, and for establishing upper bounds on minor-minimal obstructions to having treewidth at most~$k$~\cite{Lagergren98}.

The algorithm of Bodlaender and Kloks is the key subroutine in Bodlaender's linear-time algorithm for computing the treewidth of a graph~\cite{Bodlaender96}. We refer the reader to an article of the second author~\cite{Pilipczuk20a}, which gives a high-level overview of the Bodlaender-Kloks algorithm, and explains how it is applied in Bodlaender's algorithm for treewidth. (Some ideas in this exposition originate in the present article.)

\paragraph*{Our results.} The main result of this paper (Theorem~\ref{thm:main}) is that the problem of Bodlaender and Kloks can be solved by an {\em{\mso transduction}},
which is a way of describing nondeterministic transformations of relational structures using monadic second-order logic.
More precisely, we show that for every $k\in \{0,1,2,\ldots\}$ there is an \mso transduction that inputs a tree decomposition of width $k$ of a graph $G$, 
and outputs nondeterministically a tree decomposition of $G$ of optimum~width.

As a corollary of our main result, we show  (Corollary~\ref{cor:courcelle-stronger}) that an \mso transduction can compute an optimum-width tree decomposition, 
even if the input is only the graph and not a (possibly suboptimal) tree decomposition. This application is obtained by combining   
the main result of this paper with Theorem 2.4 from~\cite{bojanczyk2016definability}, which says that for every $k \in \set{0,1,2,\ldots}$ 
there is an \mso transduction  which inputs a graph of treewidth $k$ and outputs nondeterministically one of its tree decompositions of possibly suboptimal width at most $f(k)$, for some function $f$.
In particular, we thus strengthen Theorem 2.4 of~\cite{bojanczyk2016definability} by making the output a decomposition of exactly the optimum width, instead of only bounded by a function of the optimum.

Another corollary of our main result is of algorithmic nature. 
Namely, using known results for computing answers to \mso queries on structures of bounded treewidth~\cite{Bagan06,FlumFG02}, we prove the following fact.
For a fixed \mso transduction $\Ii$ and input relational structure with bounded treewidth,
we can compute some output of $\Ii$ on the input structure (or conclude that there is no output) in time linear in the sum of the sizes of the input and the output.
See Theorem~\ref{thm:implement-transduction} in Section~\ref{sec:implementing} for a formal statement.
By combining this meta-theorem with our main result we can immediately recover the algorithmic result of Bodlaender and Kloks (Theorem~\ref{thm:bodlaender-kloks}), though without explicit bounds on the
dependence of the running time on $k$.

\paragraph*{Proof techniques.} The proof of our main result is divided into a few steps. 
First, we prove a result called the {\em{Dealternation Lemma}}, which shows that there always exists an optimum-width tree decomposition 
that has bounded ``alternation'' with respect to the input suboptimal decomposition. Intuitively, small alternation is the key property allowing an optimum-width tree decomposition to be captured
by an \mso transduction or by a dynamic programming algorithm that works on the input suboptimal decomposition.
This part of the proof essentially corresponds to the machinery of typical sequences of Bodlaender and Kloks. 
However, we find the approach via alternation more intuitive and combinatorially less complicated, and we hope that it will find applications for computing other width measures.
In fact, a similar approach has very recently been used by Giannopoulou et al.~\cite{GiannopoulouPRT16} in a much simpler setting of cutwidth to give a new fixed-parameter algorithm for this graph parameter.

Next, we derive a corollary of the Dealternation Lemma called the {\em{Conflict Lemma}}, which directly prepares us to construct the \mso transduction for the Bodlaender-Kloks problem.
The Conflict Lemma is stated in purely combinatorial terms, but intuitively it shows that some optimum-width tree decomposition of the graph can be interpreted
in the given suboptimum-width tree decomposition using subtrees that cross each other in a restricted fashion, guessable in \mso.
Finally, we formalize the intuition given by the Conflict Lemma in \mso, thus constructing the \mso transduction promised in our main result.

\section{Preliminaries and statement of the main result}\label{sec:overview}

\paragraph*{Trees, forests and tree decompositions.}
Throughout this paper all graphs are undirected, unless explicitly stated.
A {\em{forest}} (which is sometimes called a {\em{rooted forest}} in other contexts) is defined to be an acyclic graph, where every connected component has one designated node called the {\em{root}}.
This naturally imposes parent--child and ancestor--descendant relations in a (rooted) forest. We use the usual tree terminology: root, leaf, child, parent, descendant and ancestor. 
We assume that every node is its own descendant, to exclude staying in the same node we use the name \emph{strict descendant}; likewise for ancestors.
For forests we often use the name \emph{node} instead of vertex. A tree is the special case of a forest that is connected and thus has one root. 
Two nodes in a forest are called {\em{siblings}} if they have a common parent, or if they are both roots. 
Note that there is no order on siblings, unlike some models of unranked forests where siblings are ordered from left to right. 

A {\em{tree decomposition}} of a graph $G$ is a pair $t=(F,\bag)$, where $F$ is a rooted forest and $\bag(\cdot)$ is a function that associates {\em{bags}} to the nodes of $F$. A bag
is a nonempty subset of vertices of $G$. We require the following two properties: (T1) whenever $uv$ is an edge of $G$, then there exists a node of~$F$ whose bag contains both $u$ and $v$; 
and (T2) for every vertex $u$ of~$G$, the set of nodes of $F$ whose bags contain $u$ is nonempty and induces a connected subtree in~$F$.
The {\em{width}} of a tree decomposition is its maximum bag size minus $1$, and the {\em{treewidth}} of a graph is the minimum  width of its tree decomposition. 
An {\em{optimum-width}} tree decomposition is one whose width is equal to the treewidth of the underlying graph.
Note that throughout this paper all tree decompositions will be rooted forests. 
This slightly diverges from the literature where usually the shape of a tree decomposition is an unrooted~tree. 

For a tree decomposition $t=(F,\bag)$ of a graph $G$, and each node $x$ of $F$, we define the following vertex sets:
\begin{itemize}
\item The {\em{adhesion}} of $x$, denoted $\adh(x)$, is equal to $\bag(x)\cap \bag(x')$, where $x'$ is the parent of $x$ in~$F$. If $x$ is a root of $F$, we define its adhesion to be empty.
\item The {\em{margin}} of $x$, denoted $\mar(x)$, is equal to $\bag(x)\setminus \adh(x)$.
\item The {\em{component}} of $x$, denoted $\comp(x)$, is the union of the margins of all the descendants of~$x$ (including $x$ itself). 
Equivalently, it is the union of the bags of all the descendants of $x$, minus the adhesion of $x$.
\end{itemize}
Whenever the tree decomposition $t$ is not clear from the context, we specify it in the subscript, i.e., we use operators $\bag_t(\cdot)$, $\adh_t(\cdot)$, $\mar_t(\cdot)$, and $\comp_t(\cdot)$.

Observe that, by property~(T2) of a tree decomposition, for every vertex of $G$ there is a unique node whose bag contains $u$, but the bag of its parent (if exists) does not contain $u$.
In other words, there is a unique node whose margin contains $u$.
Consequently, the margins of the nodes of a tree decomposition form a partition of the vertex set of the underlying graph.

\paragraph*{Relational structures and MSO\@.} Define a \emph{vocabulary} to be a finite set of \emph{relation names}, each with associated arity that is a nonnegative integer. 
A \emph{relational structure} over the vocabulary $\Sigma$ consists of a set called the \emph{universe}, and for each relation name in the vocabulary, an associated relation of the same arity over the universe. 
To describe properties of relational structures, we use logics, mainly {\em{monadic second-order logic}} (\mso for short). This logic allows quantification both over single elements of the universe and also over subsets of the universe. 
For a precise definition of \mso, see~\cite{0030804}.

We use \mso to describe properties of graphs and tree decompositions. To do this, we need to model graphs and tree decompositions as relational structures.
A graph is viewed as a relational structure, where the universe is a disjoint union of the vertex set and the edge set of a graph. There is a single binary incidence relation, 
which selects a pair $(v,e)$ whenever $v$ is a vertex and $e$ is an incident edge. 
The edges can be recovered as those elements of the universe which appear on the second coordinate of the incidence relation; the vertices can be recovered as the rest of the universe. 
For a tree decomposition of a graph $G$, the universe of the corresponding structure consists of the disjoint union of: 
the vertex set of~$G$, the edge set of~$G$, and the node set of the tree decomposition. 
There is the incidence relation between vertices and edges, as for graphs, a binary descendant relation over the nodes of the tree decomposition, 
and a binary bag relation which selects pairs $(v,x)$ such that $x$ is a node of the tree decomposition whose bag contains vertex $v$ of the graph. 
The nodes of the decomposition can be recovered as those which are their own descendants, since we assume that the descendant relation is reflexive.
Note that thus, the representation of a tree decomposition as a relational structure contains the underlying graph as a substructure.

\paragraph*{MSO transductions.}
Suppose that $\Sigma$ and $\Gamma$ are  vocabularies. 
Define a \emph{transduction} with input vocabulary $\Sigma$ and output vocabulary $\Gamma$ to be a set of pairs 
$$\textrm{(input structure over $\Sigma$, output structure over $\Gamma$)}$$ 
that is invariant under isomorphism of relational structures. When talking about transductions on graphs or tree decompositions, we use the representations described in the previous paragraph. 
Note that a transduction is a relation and not necessarily a function, thus it can have many possible outputs for the same input.
A transduction is called \emph{deterministic} if it is a partial function (up to isomorphism).
For example, the subgraph relation is a transduction from graphs to graphs, but it is not deterministic since a graph can have many subgraphs. 
On the other hand, the transformation that inputs a tree decomposition and outputs its underlying graph is a deterministic~transduction.

We use \mso transductions, which are a special case of transductions that can be defined using the logic \mso. 
The precise definition is in Section~\ref{sec:construction}, but the main idea is that an \mso transduction is a finite composition of transductions of the following types: 
copy the input a fixed number of times, nondeterministically color the universe of the input, and add new predicates to the vocabulary with interpretations given by \mso formulas over the input vocabulary.
The notion of transductions we use is borrowed from our previous work~\cite{bojanczyk2016definability} and differs syntactically from the common definition that can be found, for instance, in the book of Courcelle and Engelfriet~\cite{0030804}. However,  both definitions can be easily seen to be equivalent. We invite the reader to~\cite{0030804} for a broader discussion of the role of \mso transduction in the theory of formal languages for graphs.

\paragraph*{The main result.}
We  now state the main contribution of this paper, which  is an \mso version of the algorithm of Bodlaender and Kloks.

\begin{thm}\label{thm:main}
For every $k \in \set{0,1,2,\ldots}$ there is an  \mso transduction from tree decompositions to tree decompositions such that for every input tree decomposition $t$:
\begin{itemize}
	\item if $t$ has width at most $k$, then there is at least one output; and
	\item every output is an optimum-width tree decomposition of the underlying graph of $t$.
\end{itemize}
\end{thm}
Let us stress that the transduction of Theorem~\ref{thm:main} is not deterministic, that is,~it might have several outputs on the same input.
Using Theorem~\ref{thm:main}, we prove that an \mso transduction can compute an optimum-width tree decomposition given only the graph.
\begin{cor}\label{cor:courcelle-stronger}
For every  $k \in \set{0,1,2,\ldots}$ there is an  \mso transduction from graphs to tree decompositions such that for every input graph $G$:
\begin{itemize}
	\item if $G$ has treewidth at most $k$, then there is at least one output; and
	\item every output is a tree decomposition of $G$ of optimum width.
\end{itemize}
\end{cor}
\begin{proof}
	Theorem 2.4 of~\cite{bojanczyk2016definability} says that for every $k \in \set{0,1,2,\ldots}$ there is an \mso transduction with exactly the properties stated in the statement, 
	except that when the input has treewidth $k$, then the output tree decompositions have width at most $f(k)$, for some function $f \colon \Nat \to \Nat$. 
	By composing this transduction with the transduction given by Theorem~\ref{thm:main}, applied to $f(k)$, we obtain the claim.
\end{proof}
We remark that all the arguments that we will use in the proof of Theorem~\ref{thm:main} are constructive, hence the \mso transduction whose existence is asserted in Theorem~\ref{thm:main}
can be computed given $k$ as the input. The same holds also for the \mso transduction given by Theorem 2.4 of~\cite{bojanczyk2016definability}, even though this is not explicitly stated in this work.
As a result, the \mso transduction of Corollary~\ref{cor:courcelle-stronger} can be also computed given $k$. In order not to obfuscate the presentation with computability issues of secondary relevance 
and straightforward nature, we choose to rely on the reader in verifying these claims.

\paragraph*{Structure of the paper.} Sections~\ref{sec:dealternation}--\ref{sec:construction} are devoted to the proof of Theorem~\ref{thm:main}.
First, in Section~\ref{sec:dealternation} we formulate the Dealternation Lemma.
Its proof is deferred to Section~\ref{sec:dealternation-proof} in order not to disturb the flow of the reasoning.
Next, in Section~\ref{sec:conflict} we prove the Conflict Lemma, which is a corollary of the Dealternation Lemma.
Finally, in Section~\ref{sec:construction} we introduce formally \mso transductions and use the combinatorial insight given by the Conflict Lemma to prove Theorem~\ref{thm:main}.
In Section~\ref{sec:implementing} we show how \mso transductions can be implemented in linear fixed-parameter time on structure of bounded treewidth, and we discuss the corollaries of
combining this result with our \mso transduction for the Bodlaender-Kloks problem.
This result relies on a normalization theorem for \mso transductions, whose proof is deferred to Section~\ref{sec:normalization} due to its technicality.
Finally, in Section~\ref{sec:conc} we give some concluding remarks.

\section{Dealternation}
\label{sec:dealternation}
In this section we introduce the Dealternation Lemma, which intuitively says that for a tree decomposition $t$ of bounded, though possibly suboptimal width, 
there always exists an optimum-width decomposition in which every subtree of $t$ is broken into small number of ``pieces''.
We begin by defining factors, which is our notion of ``pieces'' of a tree decomposition.

\paragraph*{Factors and factorizations.}
Intuitively, a factor is a set  of nodes in a forest that respects the tree structure. We define three kinds of factors: tree factors, forest factors, and context factors. 
A {\em{tree factor}} in a forest is a set of nodes obtained by taking all (not necessarily strict) descendants of some node, which is called the \emph{root} of the tree factor. 
Define a  \emph{forest factor} to be a nonempty union of tree factors whose roots are siblings. 
These roots are called the  {\em{roots}} of the forest factor. In particular, a tree factor is also a forest factor, with one root.

\smallskip
\begin{center}
\begin{tabular}{ccc}
		\includegraphics[scale=0.23,page=15]{flat} &
		\includegraphics[scale=0.23,page=14]{flat}\\
	a forest factor & a context factor	
	\end{tabular}
\end{center}
\medskip

A {\em{context factor}} is the difference $X-Y$ for a tree factor $X$ and a forest factor $Y$, where the root of $X$ is a strict ancestor of every root of $Y$. For a context factor $X-Y$, its root is defined to be the root of  $X$, while the roots of $Y$ are called the {\em{appendices}}. Note that a context factor always contains a unique node that is the parent of all its appendices.

Forest factors and context factors will be jointly called {\em{factors}}. The following lemma can be proved by a straightforward case study, and hence we leave its proof to the reader.

\begin{lem}\label{lem:union-factor}
	The union of two intersecting factors in the same forest is also a factor.\end{lem}

For a subset $U$ of nodes of a forest, a $U$-factor is a factor that is entirely contained in $U$.
A {\em{factorization}} of $U$ is a partition of $U$ into $U$-factors.
A $U$-factor is {\em{maximal}} if no other $U$-factor contains it as a strict subset.

\begin{lem}\label{lem:coarsest}
Suppose $U$ is a subset of nodes of a forest. Then the maximal $U$-factors form a factorization of~$U$.
\end{lem}
\begin{proof}
Every node of $U$ is contained in some factor, e.g.,~a singleton  factor (which has forest or context type depending on whether the node is a leaf or not). 
Thus, every node of $U$ is also contained in some maximal $U$-factor. On the other hand, two different maximal $U$-factors must be disjoint, 
since otherwise by Lemma~\ref{lem:union-factor}, their  union would  also be a $U$-factor, contradicting maximality. 
\end{proof}

The set of all maximal $U$-factors will be called the {\em{maximal factorization}} of $U$, and will be denoted by $\refine(U)$. We specify the forest in the subscript whenever it is not clear from the context. Lemma~\ref{lem:coarsest} asserts that $\refine(U)$ is indeed a factorization of $U$. 
Note that the maximal factorization of $U$ is the coarsest in the following sense: in every factorization of~$U$, each of its factors is contained in some factor of $\refine(U)$.
In particular, the maximal factorization has the smallest number of factors among all factorizations of $U$.

In the sequel, we will need the following simple result about relation between the maximal factorizations of a set and of its complement.
Its proof is a part of the proof of the Dealternation Lemma, and can be found in Section~\ref{sec:factorizations-appendix} (see Lemma~\ref{lem:complement-full-bound} there).

\begin{lem}\label{lem:complement-full-bound-main}
Suppose $(U,W)$ is a partition of the node set of a rooted forest $F$, and let $k$ be the number of factors in the maximal factorization of $W$.
Then the maximal factorization of $U$ has at most $k+1$ forest factors and at most $2k-1$ context factors.
\end{lem}

\paragraph*{Elimination forests.}
The general definition of a tree decomposition is flexible and allows for multiple combinatorial adjustments.
Here, we will rely on a normalized form that we call {\em{elimination forests}}, which are essentially tree decompositions where all the margins have size exactly~$1$.
The definition of treewidth via elimination forests resembles the definition of pathwidth via the so-called {\em{vertex separation number}}~\cite{Kinnersley92}.

\begin{defi}
Suppose $G$ is a graph. An {\em{elimination forest}} of $G$ is a rooted forest $F$ on the same vertex set as $G$ such that
$G$ is contained in the ancestor-descendant closure of $F$; that is, whenever $uv$ is an edge of $G$, then $u$ is an ancestor of $v$ in $F$ or vice versa.
\end{defi}

Elimination forests are used to define the graph parameter {\em{treedepth}}, which is equal to the minimum depth of an elimination forest of a graph.
To define treewidth, we need to take a different measure than just the depth, as explained next.

Suppose $F$ is an elimination forest of $G$. Endow $F$ with the following bag function $\bag(\cdot)$.
For any vertex $u$ of $G$, assign to $u$ the bag $\bag(u)$ consisting of $u$ and all the ancestors of $u$ in $F$ that have a neighbor among the descendants of $u$ in $F$.
The following claim follows by verifying the definition of a tree decomposition; we leave the easy proof to the reader.

\begin{clm}
If $F$ is an elimination forest of $G$ and $\bag(\cdot)$ is defined as above, then $(F,\bag)$ is a tree decomposition of $G$.
Further, for every vertex $u$ of $G$, the margin of $u$ in $(F,\bag)$ is~$\{u\}$.
\end{clm}

The tree decomposition $(F,\bag)$ defined above is said to be {\em{induced}} by the elimination forest~$F$.
Observe that if $t=(F,\bag)$ is induced by~$F$, then for any vertex~$u$, the component of $u$ in~$t$ consists of all the descendants of $u$ in $F$.

\begin{figure}[htbp!]
        \centering
                \def\svgwidth{0.9\textwidth}
                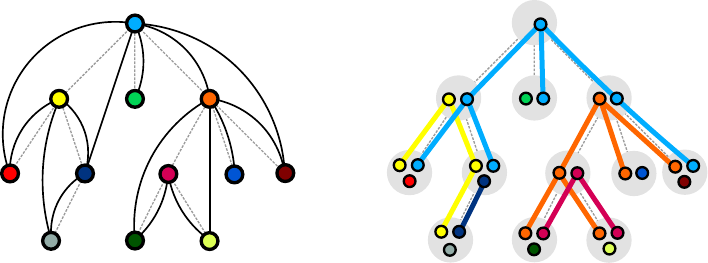
\caption{Construction of the induced tree decomposition from an elimination forest. 
The graph edges are depicted in black, the child-parent relation of the forest is depicted as dashed grey lines.}\label{fig:dispersed}
\end{figure}

One can reformulate the construction given above as follows. First, put every vertex $u$ into its bag $\bag(u)$.
Then, examine every neighbor $v$ of $u$, and if $v$ is a descendant of $u$ in $F$, then add $u$ to every bag on the path from $v$ to $u$ in $F$.
Thus, every vertex $u$ is ``smeared'' onto a subtree of $F$, where $u$ is the root of this subtree and its leaves correspond to those neighbors of $u$ that are also its descendants in $F$.
This construction is depicted in Figure~\ref{fig:dispersed}.

The {\em{width}} of an elimination forest is simply the width of the tree decomposition induced by it.
Consequently, the width of an elimination forest is never smaller than the treewidth of a graph.
The next result shows that in fact there is always an elimination forest of optimum width. The proof follows by a simple surgery on an optimum-width tree decomposition, and can be found in 
Section~\ref{sec:decompositions-appendix} (see Lemma~\ref{lem:sep-forest-opt} there). 

\begin{lem}\label{lem:sep-forest-opt-main}
For every graph $G$ there exists an elimination forest of $G$ whose width is equal to the treewidth of $G$.
\end{lem}

\paragraph*{Dealternation Lemma.} We are finally ready to state the Dealternation Lemma.

\newcommand{\dl}{D}
\newcommand{\dlref}[1]{\ref{#1}}

\begin{lem}[Dealternation Lemma]\label{lem:dealternation}
There exist functions $f(k)\in \Oh(k^3)$ and $g(k)\in \Oh(k^4)$ such that the following holds.
Suppose that $t$ is a tree decomposition of a graph $G$ of width $k$. 
Then there exists an optimum-width elimination forest $F$ of $G$ such that:
\begin{enumerate}[(\dl 1)]
	\item \label{it:few-factors} for every node $x$ of $t$, the maximal factorization $\refine_F(\comp_t(x))$ has at most $f(k)$ factors;
	\item \label{it:few-context-factors} for every node $x$ of $t$, there are at most $g(k)$ children of $x$ in the set
	\vspace{-0.1cm}
	\begin{align*}
		\set{\, y\ \colon\ \mbox{$y$ is a node of $t$ with at least one context factor in $\refine_F(\comp_t(y))$}\,}.
	\end{align*}
\end{enumerate} 
\end{lem}

Note that in the statement of the Dealternation Lemma, the vertex set of $G$ is at the same time the node set of the forest $F$.
Thus, $\refine_F(\comp_t(x))$ denotes the maximal factorization of $\comp_t(x)$, treated as a subset of nodes of $F$.

The proof of the Dealternation Lemma uses essentially the same core ideas as the correctness proof of the algorithm of Bodlaender and Kloks~\cite{BodlaenderK96}.
We include our proof for several reasons.
First, unlike in~\cite{BodlaenderK96}, in our setting we cannot assume that $t$ has binary branching, as is the case in~\cite{BodlaenderK96}.
In fact, condition~\dlref{it:few-context-factors} is superfluous when $t$ has binary branching.
Second, our formulation of the Dealternation Lemma highlights the key combinatorial property, 
which is expressed as the existence of a single elimination forest $F$ that behaves nicely with respect to the input decomposition~$t$.
This property is somehow implicit~\cite{BodlaenderK96}, where the existence of nicely-behaved optimum-width tree decompositions is argued along performing dynamic programming.
For this reason, we find the new formulation more explanatory and potentially interesting on its own.

For now we take the Dealternation Lemma for granted and we proceed with the proof of Theorem~\ref{thm:main}.
The proof of the Dealternation Lemma can be found in Section~\ref{sec:dealternation-proof}.

\section{Using the Dealternation Lemma}\label{sec:conflict}

In this section we use the Dealternation Lemma to show that an optimum-width elimination forest of a graph can be interpreted in a suboptimum-width tree decomposition.
For this, we need to develop a better understanding of the combinatorial insight provided by the Dealternation Lemma, which is expressed via an auxiliary graph, called the {\em{conflict graph}}.

Suppose $G$ is a graph, $t$ is a tree decomposition of $G$ of width $k$, and $F$ is an elimination forest of $G$.
Let $\phi$ be the mapping that sends each vertex $u$ of $G$ to the unique node of $t$ that contains $u$ in its margin.
For a vertex $u$ of $G$, we define the {\em{stain}} of $u$, denoted $S_u$, which is a subgraph of the underlying forest of $t$, as follows.
For every child $v$ of $u$ in $F$, find the unique path in $t$ between $\phi(u)$ and $\phi(v)$.
Then stain $S_u$ consists of the node $\phi(u)$ and the union of these paths.
Note that if $u$ is a leaf of $F$, then the stain $S_u$ consists only of the node~$\phi(u)$.
Define the {\em{conflict graph}} $H(t,F)$ as follows. 
The vertices of $H(t,F)$ are the vertices of~$G$, and vertices $u$ and $v$ are adjacent in $H(t,F)$ if and only their stains $S_u$ and $S_v$ have a node in common.
The main result of this section can be formulated as follows.

\begin{lem}[Conflict Lemma]\label{lem:conflict}
There is a function $h(k)\in \Oh(k^7)$ such that if $t$ and~$F$ are as in the Dealternation Lemma, then their conflict graph $H(t,F)$ admits a proper coloring with $h(k)$ colors.
\end{lem}

Recall here that a proper coloring of a graph is a coloring of its vertex set such that no two adjacent vertices receive the same color.
The rest of this section is devoted to the proof of the Conflict Lemma. From now on, we assume that $G,t,F$ are as in the Dealternation Lemma, and we denote $H=H(t,F)$.

Observe that the conflict graph $H$ is an intersection graph of a family of subtrees of a forest (here, a {\em{subtree}} of a forest $F$ is simply a connected subgraph of $F$). 
It is well-known (see, e.g.,~\cite{Golumbic-book}) that this property precisely characterizes the class of chordal graphs (graphs with no induced cycle of length larger than $3$), so $H$ is chordal.
 Chordal graphs are known to be perfect (again see, e.g.,~\cite{Golumbic-book}), hence the chromatic number of a chordal graph (the minimum number of colors needed in a proper coloring) is equal to 
the size of the largest clique in it. On the other hand, subtrees of a forest are known to satisfy the so-called Helly property: 
whenever $\mathcal{F}$ is some family of subtrees such that the subtrees in $\mathcal{F}$ pairwise intersect, then in fact there is a node of the forest that belongs to all the subtrees in $\mathcal{F}$.
This means that the largest clique in an intersection graph of a family of subtrees of a forest can be obtained by taking all the subtrees that contain some fixed node.
Therefore, to prove the Conflict Lemma it is sufficient to prove the following claim.

\begin{clm}\label{cl:load}
There exists a function $h(k)\in \Oh(k^7)$ such that every node of $t$ belongs to at most $h(k)$ of the stains $\{S_u\colon u\in V(G)\}$.
\end{clm}

In the remainder of this section we prove Claim~\ref{cl:load}.
Fix any node $x$ of $t$, and let $y_1,y_2,\ldots,y_p$ be its children in $t$.
Consider the following partition of the vertex set of $G$:
\begin{equation*}
\Pi=(\comp_t(y_1),\comp_t(y_2),\ldots,\comp_t(y_p),\mar_t(x),V(G)\setminus \comp_t(x))
\end{equation*}
Define a factorization $\Phi$ of the whole node set of $F$ as follows: for each set $X$ from the partition~$\Pi$, take its maximal factorization $\refine_F(X)$, and define $\Phi$ to be the union of these
maximal factorizations. Thus, $\Phi$ is a factorization that refines the partition $\Pi$.
Since the number of children $y_i$ is unbounded, we cannot expect that $\Phi$ has a small number of factors, but at least it has a small number of context factors.

\begin{clm}\label{cl:few-context}
Factorization $\Phi$ contains at most $g(k)\cdot f(k)+2f(k)+k$ context factors, where $f$ and $g$ are as in the Dealternation Lemma.
\end{clm}
\begin{proof}
By the Dealternation Lemma, each of the sets $\comp_t(y_1),\ldots,\comp_t(y_p),\comp_t(x)$ has at most $f(k)$ factors in its maximal factorization in $F$.
Moreover, only at most $g(k)$ of the sets $\comp_t(y_1),\ldots,\comp_t(y_p)$ can have a context factor in their maximal factorizations.
Hence, the maximal factorizations of sets $\comp_t(y_1),\ldots,\comp_t(y_p)$ introduce at most $g(k)\cdot f(k)$ context factors to the factorization $\Pi$. 
Since the maximal factorization of $\comp_t(x)$ has at most $f(k)$ factors as well, by Lemma~\ref{lem:complement-full-bound-main} we deduce that the maximal factorization of
$V(G)\setminus \comp_t(x)$ has at most $2f(k)-1$ context factors.
Finally, the cardinality of $\mar_t(x)$ is at most $k+1$, so in particular its maximal factorization has at most $k+1$ factors in total.
Summing up all these upper bounds, we conclude that $\Phi$ has at most $g(k)\cdot f(k)+2f(k)+k$ context factors.
\cqed\end{proof}

With Claim~\ref{cl:few-context} in hand, we complete now the proof of Claim~\ref{cl:load}.
Take any vertex $u$ such that $x$ belongs to the stain $S_u$.
This means that either
\begin{enumerate}[(i)]
\item\label{c:margin} $u$ belongs to the margin of $x$, or 
\item\label{c:pass} $u$ does not belong to the margin of $x$, but $u$ has a child $v$ in $F$ such that the unique path in $t$ between $\phi(u)$ and $\phi(v)$ passes through $x$.
\end{enumerate}
The number of vertices $u$ satisfying~\ref{c:margin} is bounded by the size of the margin of $x$, which is at most $k+1$, hence we focus on vertices $u$ that satisfy~\ref{c:pass}.
Observe that condition~\ref{c:pass} in particular means that $u$ and $v$ belong to different parts of partition $\Pi$, so also to different factors of the factorization $\Phi$.
Since $u$ is the parent of $v$ in $F$, this means that the unique factor of~$\Phi$ that contains $u$ must be a context factor, and $u$ must be the parent of its appendices.
Consequently, the number of vertices $u$ satisfying~\ref{c:pass} is upper bounded by the number of context factors in factorization $\Phi$, which is at most $g(k)\cdot f(k)+2f(k)+k$
by Claim~\ref{cl:few-context}. We conclude that the number of stains $S_u$ containing $x$ is at most
\begin{equation*}
h(k):=g(k)\cdot f(k)+2f(k)+2k+1;
\end{equation*}
in particular $h(k)\in \Oh(k^7)$. This concludes the proof of Claim~\ref{cl:load}, so also the proof of the Conflict Lemma is complete.

\section{Constructing the transduction}\label{sec:construction}

We now use the understanding gathered in the previous sections to give an \mso transduction that takes a tree decomposition of a graph of suboptimum width, and produces an optimum-width tree decomposition.
First, we need to precisely define \mso transductions.

\paragraph*{\bfseries{\scshape{MSO}} transductions.} 
Formally, an \mso transduction is any transduction that can be obtained by composing a finite number of transductions of the following kinds.
Note that kind 1 is a partial function, kinds 2, 3, 4 are functions, and kind 5 is a relation.
\begin{enumerate}
	\item {\bf Filtering.} For every \mso sentence $\varphi$ over the input vocabulary there is transduction that filters out structures where $\varphi$ is satisfied. Formally, the transduction is the partial identity whose domain consists of the structures that satisfy the sentence. The input and output vocabularies are the same.
	\item {\bf Universe restriction.} For every \mso formula $\varphi(x)$ over the input vocabulary with one free first-order variable there is a transduction, which restricts the universe to those elements that satisfy $\varphi$. The input and output vocabularies are the same, the interpretation of each relation in the output structure is defined as the restriction of its interpretation 
	in the input structure to tuples of elements that remain in the universe.
	\item {\bfseries{\scshape{MSO}} interpretation.} This kind of transduction changes the vocabulary of the structure while keeping the universe intact. For every relation name $R$ of the output vocabulary, there is an \mso formula $\varphi_R(x_1,\ldots,x_k)$ over the input vocabulary which has as many free first-order variables as the arity of $R$. The output structure is obtained from the input structure by keeping the same universe, and interpreting each relation $R$ of the output vocabulary as the set of those tuples $(x_1,\ldots,x_k)$ that satisfy $\varphi_R$.
	\item {\bf Copying.} For  $k \in \set{1,2,\ldots}$, define $k$-copying to be the transduction which inputs a structure and outputs a structure consisting of $k$ disjoint copies of the input.
	Precisely, the output universe consists of $k$ copies of the input universe.
	The output vocabulary is the input vocabulary enriched with a binary predicate $\mathsf{copy}$ that selects copies of the same element, and unary predicates $\mathsf{layer}_1,\mathsf{layer}_2,\ldots,\mathsf{layer}_k$ which select elements belonging to the first, second, etc.\ copies of the universe.
	In the output structure, a relation name $R$ of the input vocabulary is interpreted as the set of all those tuples over the output structure, where the original elements of the copies were in relation $R$
	in the input structure.
	\item {\bf Coloring.} We add a new unary predicate to the input structure. Precisely, the universe as well as the interpretations of all relation names of the input vocabulary stay intact,
	but the output vocabulary has one more unary predicate. For every possible interpretation of this unary predicate, there is a different output with this interpretation implemented.
\end{enumerate}
We remark that the above definition is easily equivalent to the one used in~\cite{bojanczyk2016definability}, where filtering, universe restriction, and \mso interpretation are merged into one kind of a transduction.

\paragraph*{Proving the main result.} We are finally ready to prove our main result, Theorem~\ref{thm:main}.
The proof is broken down into several steps. The first, main step shows that an \mso transduction can output optimum-width elimination forests.
Here, an elimination forest of a graph $G$ is encoded by enriching the relational structure encoding $G$ with a single binary relation interpreted as the child relation of $F$.
Note that the definition of an elimination forest is \mso-expressible: there is an \mso sentence that checks whether the additional relation indeed encodes an elimination forest of the graph.

\begin{lem}\label{lem:step-main}
For every $k \in \set{0,1,2,\ldots}$, there is an \mso transduction from tree decompositions to elimination forests such that for every input tree decomposition $t$:
\begin{itemize}
\item every output is an elimination forest of the underlying graph of $t$; and
\item if $t$ has width at most $k$, then there is at least one output that is an elimination forest of optimum width.
\end{itemize}
\end{lem}
\begin{proof}
Observe that the verification whether the width of $t$ is at most $k$ can be expressed by an \mso sentence, so we can first use filtering to filter out any input tree decomposition $t$ whose width is larger than $k$;
for such decompositions, the transduction produces no output.
Let $G$ be the underlying graph of $t$, and let $\phi$ be the mapping that sends each vertex $u$ of $G$ to the unique node of $t$ whose margin contains $u$.
By the Conflict Lemma, there exists some elimination forest $F$ of $G$ of optimum width such that the conflict graph $H(t,F)$ admits some proper coloring $\lambda$ with $h(k)$ colors.
The constructed \mso transduction attempts at guessing and interpreting $F$ as follows.

First, using coloring and filtering, we guess the coloring $\lambda$, represented as a partition of the vertex set of $G$.
Then, again using coloring and filtering, for every vertex $u$ of $G$ we guess whether $u$ is a root of $F$, and if not, then we guess the color under $\lambda$ of the parent of $u$ in $F$.

Next, for every color $c$ used in $\lambda$, we guess the forest
\begin{equation*}
M_c:=\bigcup_{u\in \lambda^{-1}(c)} S_u,
\end{equation*}
where $S_u$ is the stain of $u$ in $t$, defined as in Section~\ref{sec:conflict} for the elimination forest $F$. 
Note that the stains $\{S_u\colon u\in \lambda^{-1}(c)\}$ are pairwise disjoint, because $\lambda$ is a proper coloring of the conflict graph $H(t,F)$.
Thus, the connected components of $M_c$ are exactly these stains.
Observe also that $M_c$ is a subgraph of the decomposition $t$, so we can emulate guessing $M_c$ in an \mso transduction working over $t$ by guessing 
the subset of those nodes of $t$, for which the edge of $t$ connecting the node and its parent belongs to $M_c$.

Having done all these guesses, we can interpret the child relation of $F$ using an \mso predicate as follows.
Fix a pair of vertices $u$ and $v$, and let $c$ be the guessed color of $u$ under~$\lambda$.
Then one can readily check that $u$ is the parent of $v$ in $F$ if and only if the following conditions are satisfied:
\begin{itemize}
\item we have guessed that $v$ is not a root of $F$, 
\item we have guessed that the color of the parent of $v$ in $F$ is $c$, and
\item $u$ is the unique vertex of color $c$ such that $\phi(u)$ belongs to the same connected component of $M_c$ as $\phi(v)$.
\end{itemize}
It can be easily seen that these conditions can be expressed by an \mso formula with two free variables $u$ and $v$.

Finally, we filter out all the wrong guesses by verifying, using an \mso sentence, whether the interpreted child relation on the vertices of $G$ indeed forms a rooted forest, 
and whether this forest is an elimination forest of $G$. Obviously, the elimination forest $F$ was obtained for at least one of the guesses, and survives this filtering.
At the end, we remove the nodes of decomposition $t$ from the structure using universe restriction.
\end{proof}

Next, we need to construct the induced tree decomposition out of an elimination forest.

\begin{lem}\label{lem:step-induce}
There is an \mso transduction from elimination forests to tree decompositions that on each input elimination forest has exactly one output, which is the tree decomposition induced by the input.
\end{lem}
\begin{proof}
We copy the vertex set of the graph two times, and declare the second copies to be the nodes of the constructed tree decomposition. 
Using the child relation of the input elimination forest, we can interpret in \mso the descendant relation
in the forest of the decomposition. Finally, the bag relation in the induced tree decomposition, as defined in Section~\ref{sec:dealternation}, can be easily interpreted using an \mso formula. 
\end{proof}

Finally, so far the transduction can output tree decompositions of suboptimal width, which should be filtered out.
For this, we need the following \mso-expressible predicate.

\begin{lem}\label{lem:step-filter}
For every $k \in \set{0,1,2,\ldots}$, there is an \mso-sentence over tree decompositions that holds if and only if the given tree decomposition has width at most $k$ and its width is optimum for the underlying graph.
\end{lem}
\begin{proof}
Let $t$ be the given tree decomposition of a graph $G$.
Obviously, we can verify using an \mso sentence whether the width of $t$ is at most $k$.
To check that the width of $t$ is optimum, we could use the fact that graphs of treewidth $k$ are characterized by a finite list of forbidden minors, but we choose to apply the following different strategy.
Let $R_k$ be the \mso transduction that is the composition of the transductions of Lemmas~\ref{lem:step-main} (for parameter $k$) and~\ref{lem:step-induce}.
Provided the input tree decomposition $t$ has width at most $k$, transduction $R_k$ outputs some set of tree decompositions of $G$ among which one has optimum width.
Hence, $t$ has optimum width if and only if the output $R_k(t)$ does not contain any tree decomposition of width smaller than $t$.

The Backwards Translation Theorem for \mso transductions~\cite{0030804} (see also~\cite{bojanczyk2016definability}) states that whenever $\mathcal{I}$ is an \mso transduction 
and $\psi$ is an \mso sentence over the output vocabulary,
then the set of structures on which $\mathcal{I}$ outputs at least one structure satisfying $\psi$, is \mso-definable over the input vocabulary.
Hence, for every $p<k$, there exists an \mso sentence $\varphi_{p}$ that verifies whether $R_k(t)$ outputs at least one tree decomposition of width at most $p$.
Therefore, we can check whether $t$ has optimum width by making a disjunction over all $\ell$ with $0\leq \ell\leq k$ of the
sentences stating that $t$ has width exactly $\ell$ and $R_k(t)$ does not output any tree decomposition of width less than $\ell$.
\end{proof}

Theorem~\ref{thm:main} now follows by composing the \mso transductions given by Lemmas~\ref{lem:step-main} and~\ref{lem:step-induce}, 
and at the end applying filtering using the predicate given by Lemma~\ref{lem:step-filter}.

\section{Implementing \mso transductions in FPT time}\label{sec:implementing}

In this section we prove that \mso transductions on relational structures of bounded treewidth can be implemented in linear fixed-parameter time.
To state this result formally, we first need to introduce some definitions regarding measuring the input and output size of the algorithm.

In the following, by the {\em{size}} of an \mso transduction $\Ii$, denoted $\|\Ii\|$, we mean the sum of sizes of its atomic transductions.
Here, the size of a copying step is the number of copies it produces, the size of a coloring step is $1$, and the size of a transduction of any other type is the total size of \mso formulas involved in its description.

By the treewidth of a relational structure we mean the treewidth of its Gaifman graph; that is, a graph whose vertices are elements of the structure, and
two elements are adjacent if and only if they appear together in some tuple of some relation. The {\em{size}} of a relational structure $\As=(U,R_1,R_2,\ldots,R_c)$, where $U$ is the 
universe and $R_i$ is a relation of arity $r_i$, for $i=1,\ldots,c$, is defined as
$$\|\As\|=|U|+\sum_{i=1}^c r_i\cdot |R_i|.$$
We say that an algorithm that receives a structure $\As$ on input {\em{implements $\Ii$ on $\As$}} if it either correctly concludes that $\Ii(\As)$ is empty, or outputs an arbitrary structure belonging to $\Ii(\As)$.

We may now formally state the algorithmic result for \mso transductions..

\begin{thm}\label{thm:implement-transduction}
There is an algorithm that, given an \mso transduction $\Ii$ and a relational structure $\As$ over the input vocabulary of $\Ii$, implements $\Ii$ on $\As$ in time 
$f(\|\Ii\|,w)\cdot (n+m)$, where $n$ and $w$ are the size and the treewidth of the input structure, respectively,
$m$ is the size of the output structure (or $0$ if $\Ii(\As)$ is empty), and $f$ is a computable function.
\end{thm}

The cornerstone of the proof of Theorem~\ref{thm:implement-transduction} is a normalization theorem for \mso transductions: every \mso transduction can be written in a simple normal form
that allows for algorithmic treatment.
To describe this form, it will be useful to introduce another type of an \mso transduction, which is a special case of interpretation.
By a {\em{renaming}} we mean an interpretation step that only renames symbols from the signature, possibly dropping some of them.
Precisely, if the input vocabulary is $\Sigma$ and the output vocabulary is $\Gamma$, then there is an injective function $\rho\colon \Gamma\to \Sigma$ such
each symbol $R\in \Gamma$, say of arity $r$,
is interpreted by the formula $\phi_R(x_1,\ldots,x_r)=\rho(R)(x_1,\ldots,x_r)$. We can now state the normalization theorem.

\begin{thm}[Normal form for \mso transductions]\label{thm:mso-normalization}
Suppose $\Ii$ is an \mso transduction. Then $\Ii$ may be represented in the form
$$\Ii= \Ii_{\mathrm{rename}}\circ \Ii_{\mathrm{restrict}}\circ\Ii_{\mathrm{interprete}}\circ \Ii_{\mathrm{copy}}\circ \Ii_{\mathrm{filter}}\circ \Ii_{\mathrm{color}},$$
where the above are \mso transductions as follows:
\begin{itemize}
\item $\Ii_{\mathrm{color}}$ is a finite sequence of coloring steps;
\item $\Ii_{\mathrm{filtering}}$ is a single filtering step;
\item $\Ii_{\mathrm{copy}}$ is a single copying step;
\item $\Ii_{\mathrm{interprete}}$ is a single interpretation step;
\item $\Ii_{\mathrm{restrict}}$ is a single universe restriction step;
\item $\Ii_{\mathrm{rename}}$ is a single renaming step.
\end{itemize}
Moreover, there is an algorithm that, given $\Ii$, computes the normal form as above.
\end{thm}

The proof of Theorem~\ref{thm:mso-normalization} roughly proceeds as follows. We write the given \mso transduction as a sequence of atomic transductions, each being a coloring, filtering, copying, interpretation,
or universe restriction step. Then, we give a number of swapping and merging rules that enable us to swap these transductions while modifying them slightly. It is shown that by applying the rules
exhaustively, we eventually arrive at the claimed normal form. While basically all the rules are straightforward, their full verification takes some effort. We give the proof of Theorem~\ref{thm:mso-normalization}
in Section~\ref{sec:normalization} for completeness, while for now let us take it for granted and proceed with the proof of Theorem~\ref{thm:implement-transduction}.

\begin{proof}[Proof of Theorem~\ref{thm:implement-transduction}]
By Theorem~\ref{thm:mso-normalization}, we can assume that $\Ii$ is in the normal form
$$\Ii= \Ii_{\mathrm{rename}}\circ \Ii_{\mathrm{restrict}}\circ \Ii_{\mathrm{interprete}}\circ \Ii_{\mathrm{copy}}\circ \Ii_{\mathrm{filter}}\circ \Ii_{\mathrm{color}}.$$
Suppose further that $\Ii_{\mathrm{color}}$ is a sequence of coloring steps that introduce new unary predicates $X_1,X_2,\ldots,X_c$, for some constant $c$,
while $\Ii_{\mathrm{copy}}$ copies the universe $\ell$ times, for some constant~$\ell$.
The proof will follow from the following two claims. 
In the following we use $f$ for an arbitrary computable function, possibly different in each context.

\begin{clm}\label{cl:guess-colors}
One can in time $f(\|\Ii\|,w)\cdot n$ determine a sequence of subsets $X_1,\ldots,X_c$ of elements of $\As$ 
such that filtering $\Ii_{\mathrm{filter}}$ preserves $\As$ enriched with $X_1,X_2,\ldots,X_c$ as unary predicates, or correctly conclude that such a sequence does not exist.
\end{clm}

\begin{clm}\label{cl:interpret}
Given $\As$ enriched with unary predicates $X_1,X_2,\ldots,X_c$, one can in time $f(\|\Ii\|,w)\cdot (n+m)$ compute the output of  
$\Ii_{\mathrm{rename}}\circ \Ii_{\mathrm{restrict}}\circ \Ii_{\mathrm{interprete}}\circ \Ii_{\mathrm{copy}}$ on this structure,
where $m$ is the size of the output.
\end{clm}

Note here that in Claim~\ref{cl:interpret}, the transduction $\Ii_{\mathrm{rename}}\circ \Ii_{\mathrm{restrict}}\circ \Ii_{\mathrm{interprete}}\circ \Ii_{\mathrm{copy}}$ uses neither copying nor filtering, 
hence every input structure is mapped to exactly one output structure.

Observe that the proof follows trivially from combining Claims~\ref{cl:guess-colors} and~\ref{cl:interpret} as follows.
First, using the algorithm of Claims~\ref{cl:guess-colors} one tries to compute any sequence of element subsets $X_1,X_2,\ldots,X_c$ for which the filtering step $\Ii_{\mathrm{filter}}$ passes.
If this cannot be done, then $\Ii(\As)$ is empty, and this conclusion can be reported.
Otherwise, we plug the obtained sequence to the algorithm of Claim~\ref{cl:interpret}, thus computing an arbitrary structure from $\Ii(\As)$.

We now prove Claims~\ref{cl:guess-colors} and~\ref{cl:interpret} in order. For this, we use the following results on answering \mso queries on structures of bounded treewidth.
Suppose we are given a relational structure $\As$ with $\tw(\As)=w$. Suppose further that $\varphi(X_1,\ldots,X_c,x_1,\ldots,x_d)$ is an \mso formula over the vocabulary of
$\As$, where $X_i$ are monadic variable and $x_i$ are first-order variables. A tuple $\bar{y}=(A_1,\ldots,A_c,a_1,\ldots,a_d)$ is an {\em{answer}} to the {\em{\mso query}} $\varphi$ if
$\As\models \varphi(A_1,\ldots,A_c,a_1,\ldots,a_d)$. Flum et al.~\cite{FlumFG02} gave an algorithm that in time $f(w,\|\varphi\|)\cdot (n+m)$ outputs all the answers to $\varphi$ on $\As$, where
$n$ is the size of the universe of $\As$, $m$ is the total size of the output, and $f$ is a computable function.
Later, Bagan~\cite{Bagan06} gave an enumeration algorithm for solving \mso queries on structures of bounded treewidth: this algorithm uses $f(w,\|\varphi\|)\cdot n$ preprocessing time, 
and then reports answers to the query with delay between two consecutive reports bounded by $f(w,\|\varphi\|)\cdot |\bar{y}|$, where $|\bar{y}|$ is the size of the next answer.
A different proof of this result, but for queries using only first-order variables, was later given by Kazana and Segoufin~\cite{KazanaS13}.

In the sequel, $f$ always denotes some computable function, possibly different in each context.

\begin{proof}[Proof of Claim~\ref{cl:guess-colors}]
Let $\psi$ be the \mso formula used in the filtering step $\Ii_{\mathrm{filter}}$, which works over the input structure $\As$ enriched with sets $X_1,\ldots,X_c$.
That is, the filtering step passes only if the sets $X_1,\ldots,X_c$ guessed by $\Ii_{\mathrm{color}}$ satisfy $\As,X_1,\ldots,X_c\models \psi$.
Interpret $\psi$ as an \mso query on $t$ with free monadic variables $X_1,\ldots,X_c$.
Run the algorithm of Bagan~\cite{Bagan06} on it to enumerate only the first answer, or to conclude that there are no answers; either of these outcomes may be then reported.
The preprocessing step takes time $f(\|\psi\|,w)\cdot n$, whereas the construction of the first answer also takes time $f(\|\psi\|,w)\cdot n$, since the size of the answer is trivially bounded by $cn$.
Since $\|\psi\|\leq \|\Ii\|$, the claimed running time follows.
\cqed\end{proof}

\begin{proof}[Proof of Claim~\ref{cl:interpret}]
First, the step $\Ii_{\mathrm{copy}}$ can be just performed in time $f(\|\Ii\|,w)\cdot n$, since $\ell$ is a constant bounded in terms of~$\|\Ii\|$.
Observe here that since $\tw(\As)\leq w$, the treewidth of the structure output by $\Ii_{\mathrm{copy}}$ is bounded by $\ell(w+1)$.
This follows by replacing, in every bag of an optimum-width tree decomposition of the Gaifman graph of $\As$,
each element of the original structure with its $\ell$ copies in $\As'$, the structure output by $\Ii_{\mathrm{copy}}$.
Next, we implement $\Ii_{\mathrm{rename}}\circ \Ii_{\mathrm{restrict}}\circ \Ii_{\mathrm{interprete}}$ on $\As'$ in time $f(\|\Ii\|,w)\cdot (n+m)$ in one shot.

Take, any relation $R$ of the output vocabulary, say of arity $r$, and let $R'$ be the relation from which $R$ originates in the renaming step $\Ii_{\mathrm{rename}}$.
Let $\varphi_{R'}(x_1,\ldots,x_r)$ be the formula used in $\Ii_{\mathrm{interprete}}$ to interprete $R'$, and let $\varphi(u)$ be the formula used in $\Ii_{\mathrm{restrict}}$ to restrict the universe.
Moreover, let $\varphi'(u)$ be a formula constructed from $\varphi(u)$ by replacing every relation atom $Q(x_1,\ldots,x_q)$ by its interpretation $\varphi_Q(x_1,\ldots,x_q)$ under $\Ii_{\mathrm{interprete}}$.
Consider the formula 
$$\alpha_{R}(x_1,\ldots,x_r) = \varphi_{R'}(x_1,\ldots,x_r)\wedge \bigwedge_{i=1}^r \varphi'(x_i).$$
Observe that $\alpha_{R}(x_1,\ldots,x_r)$ in the structure $\As'$ selects exactly those tuples $(x_1,\ldots,x_r)$ that satisfy $R(x_1,\ldots,x_r)$ in the output structure.

Hence, given the structure $\As'$, we implement $\Ii_{\mathrm{rename}}\circ \Ii_{\mathrm{restrict}}\circ \Ii_{\mathrm{interprete}}$ as follows.
First, $\varphi'(u)$ can be regarded as an \mso query with one free first-order variable over $\As'$; obviously, the number of answers to this query is bounded by the size of the universe of $\As'$, which is $\ell n$.
Hence the algorithm of Flum et al.~\cite{FlumFG02} can output all the answers to this query, which are exactly the elements that are preserved in the universe by $\Ii_{\mathrm{restrict}}$, in time 
$f(\|\psi\|,\ell(w+1))\cdot h(\|\Ii\|)\cdot n$, which is bounded by $g(\|\Ii\|,w)\cdot n$ for some computable $g$. Thus, we have computed the universe of the output structure.

To compute the relations in the output structure, for every relation $R$ of the output vocabulary, say of arity $r$, apply the algorithm of Flum et al.~\cite{FlumFG02} for
the query $\alpha_R(x_1,\ldots,x_r)$ on $\As'$. Thus we compute the set of tuples selected by $R$ in the output structure in time $f(\|\Ii\|,\ell(w+1))\cdot (h(\|\Ii\|)\cdot n+m_R)$, where $m_R$ is the size 
of relation $R$ in the output.
By summing this bound through all relations of the output vocabulary, we obtain a running time of the form $g(\|\Ii\|,w)\cdot (n+m)$ for some computable $g$, where $m$ is the output size.
\cqed\end{proof}

As argued before, the proof of Theorem follows from Claims~\ref{cl:guess-colors} and~\ref{cl:interpret}.
\end{proof}

An observant reader might wonder why in the proof of Theorem~\ref{thm:implement-transduction} we actually needed the normal form provided by Theorem~\ref{thm:mso-normalization}, as it would be natural to just implement consecutive atomic transductions comprising the input transduction $\Ii$ one by one, each in linear time. There are two reasons for this. First, every coloring step introduces a large number of possible intermediate outputs, and it can happen that only few of them eventually lead to producing an output of the whole transduction $\Ii$, due to later filtering steps. At the moment of applying a coloring step it is difficult to determine which intermediate outputs will eventually get filtered out; the normal form facilitates this verification through Claim~\ref{cl:guess-colors}. Second, applying atomic transductions comprising $\Ii$ one by one and computing each intermediate output means that the running time is linear in the maximum size among the intermediate outputs. This can be much larger than the maximum among the sizes of the input and the final output, which is the measure promised in Theorem~\ref{thm:implement-transduction}. The normal form helps here in compressing a possibly long sequence of atomic transductions into a sequence manageable in its entirety.

\medskip

We now show how the result of Bodlaender and Kloks~\cite{BodlaenderK96} may be obtained as a direct corollary of our meta-results. 

\begin{cor}\label{thm:bodlaender-kloks-our}
For every $k \in \set{0,1,2,\ldots}$ there exists a linear-time algorithm that, 
given a graph $G$ and its tree decomposition of width $k$, returns a tree decomposition of $G$ of optimum width.
\end{cor}
\begin{proof}
Let $\As$ be the relational structure representing $G$ together with the input tree decomposition $t$ of $G$ of width at most $k$.
It can be easily seen that the Gaifman graph of $\As$ has treewidth at most $2k+3$, and its tree decomposition of such width can be constructed from $t$ in linear time.
To obtain a tree decomposition of $G$ of optimum width, it suffices to apply the algorithm of Theorem~\ref{thm:implement-transduction} to $\As$ and the transduction given by Theorem~\ref{thm:main}. 
For the running time bound, observe that the size of the output is bounded linearly in the size of the input.
\end{proof}

As we argued in Section~\ref{sec:overview}, our proof of Theorem~\ref{thm:main} is actually constructive: given $k$, one can compute the transduction given by Theorem~\ref{thm:main} for this value.
Thus, we can infer a slightly stronger uniform variant of Corollary~\ref{thm:bodlaender-kloks-our}, where $k$ is also given in the input and the algorithm works in linear fixed-parameter time,
that is, in time $f(k)\cdot n$ for some computable $f$, where $n$ is the size of the input.
While the uniformity of the algorithm follows from our arguments in this way, we unfortunately do not see an easy way to recover upper bounds on the running time similar to those in Theorem~\ref{thm:bodlaender-kloks} 
using our approach.

A careful reader may have observed that our claim of recovering the algorithm of Bodlaender and Kloks~\cite{BodlaenderK96} via meta-tools might seem like cheating.
Namely, the algorithms of Flum et al.~\cite{FlumFG02} and of Bagan~\cite{Bagan06}, which are invoked in the algorithm of Theorem~\ref{thm:implement-transduction}, actually use the linear-time algorithm
of Bodlaender~\cite{Bodlaender96} to compute a tree decomposition of the given structure. This algorithm, on the other hand, uses the algorithm of Bodlaender and Kloks~\cite{BodlaenderK96}
as a subroutine, thus creating a cycle of dependencies. This issue is, however, not really problematic. Namely, the algorithms of~\cite{FlumFG02,Bagan06} use the linear-time algorithm of Bodlaender~\cite{Bodlaender96}
only as an opening step, to compute a tree decomposition that will be used in further computations. In our setting, we have a tree decomposition of the input structure in our hand, so there is no need of
performing this step. Thus, we indeed obtain a new implementation of the algorithm of Bodlaender and Kloks~\cite{BodlaenderK96}. 

Note, however,
that this vicious cycle of dependencies would persist if we tried to combine Theorem~\ref{thm:implement-transduction} with the transduction of Corollary~\ref{cor:courcelle-stronger} in order to obtain
a new implementation of the linear-time algorithm of Bodlaender~\cite{Bodlaender96}.
This is because in the setting of Corollary~\ref{cor:courcelle-stronger}, there is no tree decomposition given on the input. 
Therefore, we do not obtain a new implementation of the algorithm of Bodlaender~\cite{Bodlaender96} via our meta-techniques.
We see, however, a potential for our tools to be useful in computing other types of tree-like decompositions of graphs; we discuss this matter in more details in Section~\ref{sec:conc}.

\section{Proof of the Dealternation Lemma}\label{sec:dealternation-proof}

In this section we prove the Dealternation Lemma (Lemma~\ref{lem:dealternation}), as well as some auxiliary simple facts whose proofs were omitted in Section~\ref{sec:dealternation}.
We begin with introducing some auxiliary tools on dealternation in words, as well as we give a few useful properties of maximal factorizations; in particular we prove Lemma~\ref{lem:complement-full-bound-main}.
Then, we move to elimination forests: we prove Lemma~\ref{lem:sep-forest-opt-main}, and we investigate a normalized form of elimination forests that we call {\em{reduced}}. 
Finally, we complete the proof of the Dealternation Lemma using the gathered tools.
We first show how the Dealternation Lemma follows from an auxiliary result, called Local Dealternation Lemma, which can be thought of as one ``fixing step''.
Then we conclude by proving the Local Dealternation Lemma.

\subsection{Words and alternation}\label{sec:words}

We now give some auxiliary combinatorial tools on reshuffling on a word over alphabet $\{-,+\}$ in order to reduce its ``alternation'', while preserving some extremal properties.
These results hold the essence of the technique of {\em{typical sequences}}, used by Bodlaender and Kloks in~\cite{BodlaenderK96}.

Fix the alphabet $\Sigma=\{-,+\}$. For a word $w\in \Sigma^\star$, we define:
\begin{itemize}
\item the {\em{sum of $w$}}, denoted $\wsum(w)$, is the number of $+$ in $w$, minus the number of $-$ in $w$;
\item the {\em{prefix maximum of $w$}}, denoted $\pmax(w)$, is the maximum of $\wsum(u)$ for $u$ ranging over the prefixes of $w$;
\item the {\em{prefix minimum of $w$}}, denoted $\pmin(w)$, is the minimum of $\wsum(u)$ for $u$ ranging over the prefixes of $w$.
\end{itemize}
Suppose a word $w\in \Sigma^\star$ has every position colored with some color drawn from some set of colors; in such a case, we will talk about a {\em{colored word}}.
A {\em{block}} in a colored word $w$ is a maximal set of consecutive letters colored with the same color.

We say that a colored word $w'$ is a {\em{block-shuffle}} of $w$ if $w'$ can be obtained from $w$ by permuting its letters (and keeping their colors) in such a manner that 
(i) within each color, the order of the letters remains the same as in $w$, and (ii) every block of $w$ remains contiguous in $w'$. Note that (ii) is equivalent to saying that
every block of $w$, after applying the permutation, is contained in a block of $w'$. It is clear that if $w_1$ is a block-shuffle of $w_2$, which in turn is a block-shuffle of $w_3$,
then $w_1$ is a block-shuffle of $w_3$.

\begin{figure}[h]
\centering
\begin{tikzpicture}[scale=0.68]
 \tikzstyle{vertex}=[circle,fill=black,minimum size=0.1cm,inner sep=0pt]

\begin{scope}[shift={(-5.5,0)}]
\draw[gray, dashed] (-7.5,1.6) -- (1.5,1.6);

\node[vertex] (p1) at (-5.7,1.6) {};
\node[red,thick] at (-5.4,0.5) {$+$};
\node[vertex] (p2) at (-5.1,2.2) {};
\draw[red,thick] (p1) -- (p2);

\node[red,thick] at (-4.8,0.5) {$-$};
\node[vertex] (p3) at (-4.5,1.6) {};
\draw[red,thick] (p2) -- (p3);

\node[blue,thick] at (-4.2,0.5) {$+$};
\node[vertex] (p4) at (-3.9,2.2) {};
\draw[blue,thick] (p3) -- (p4);

\node[blue,thick] at (-3.6,0.5) {$+$};
\node[vertex] (p5) at (-3.3,2.8) {};
\draw[blue,thick] (p4) -- (p5);

\node[blue,thick] at (-3.0,0.5) {$-$};
\node[vertex] (p6) at (-2.7,2.2) {};
\draw[blue,thick] (p5) -- (p6);

\node[red,thick] at (-2.4,0.5) {$-$};
\node[vertex] (p7) at (-2.1,1.6) {};
\draw[red,thick] (p6) -- (p7);

\node[blue,thick] at (-1.8,0.5) {$-$};
\node[vertex] (p8) at (-1.5,1.0) {};
\draw[blue,thick] (p7) -- (p8);

\node[red,thick] at (-1.2,0.5) {$+$};
\node[vertex] (p9) at (-0.9,1.6) {};
\draw[red,thick] (p8) -- (p9);

\node[red,thick] at (-0.6,0.5) {$+$};
\node[vertex] (p10) at (-0.3,2.2) {};
\draw[red,thick] (p9) -- (p10);

\draw[blue] (-4.45,0.8) rectangle (-2.75,0.2);
\draw[red] (-2.65,0.8) rectangle (-2.15,0.2);

\end{scope}

\begin{scope}[shift={(5.5,0)}]
\draw[gray, dashed] (-7.5,1.6) -- (1.5,1.6);

\node[vertex] (p1) at (-5.7,1.6) {};
\node[red,thick] at (-5.4,0.5) {$+$};
\node[vertex] (p2) at (-5.1,2.2) {};
\draw[red,thick] (p1) -- (p2);

\node[red,thick] at (-4.8,0.5) {$-$};
\node[vertex] (p3) at (-4.5,1.6) {};
\draw[red,thick] (p2) -- (p3);

\node[red,thick] at (-4.2,0.5) {$-$};
\node[vertex] (p4) at (-3.9,1.0) {};
\draw[red,thick] (p3) -- (p4);

\node[blue,thick] at (-3.6,0.5) {$+$};
\node[vertex] (p5) at (-3.3,1.6) {};
\draw[blue,thick] (p4) -- (p5);

\node[blue,thick] at (-3.0,0.5) {$+$};
\node[vertex] (p6) at (-2.7,2.2) {};
\draw[blue,thick] (p5) -- (p6);

\node[blue,thick] at (-2.4,0.5) {$-$};
\node[vertex] (p7) at (-2.1,1.6) {};
\draw[blue,thick] (p6) -- (p7);

\node[blue,thick] at (-1.8,0.5) {$-$};
\node[vertex] (p8) at (-1.5,1.0) {};
\draw[blue,thick] (p7) -- (p8);

\node[red,thick] at (-1.2,0.5) {$+$};
\node[vertex] (p9) at (-0.9,1.6) {};
\draw[red,thick] (p8) -- (p9);

\node[red,thick] at (-0.6,0.5) {$+$};
\node[vertex] (p10) at (-0.3,2.2) {};
\draw[red,thick] (p9) -- (p10);

\draw[red] (-4.45,0.8) rectangle (-3.95,0.2);
\draw[blue] (-3.85,0.8) rectangle (-2.15,0.2);

\end{scope}

\end{tikzpicture}
\caption{Left panel: A bichromatic word with sum $1$, prefix maximum $2$, and prefix minimum $-1$. The number of blocks is $5$. 
Right panel: The word after swapping the second and the third block, indicated by blue and red box, respectively. Note that this swapping operation satisfies the prerequisites of Claim~\ref{cl:swap},
and consequently the prefix maximum of the word does not increase.}
\label{fig:swap}
\end{figure}

Informally, the main result of this section can be stated as follows: 
provided a colored word $w$ has bounded prefix maximum, and prefix minima within colors are also not too small,
then there exists a block-shuffle of $w$ that achieves a small number of blocks of each color.
The formal statement follows.

\begin{lem}\label{lem:reorder-two}
Suppose $w\in \Sigma^\star$ is colored with two colors.
Suppose further that $\pmax(w)\leq a$ for some nonnegative integer $a$, and if $u$ is a word derived from $w$ by restricting it to all the letters of one of the colors, then $\pmin(u)\geq -b$, for some nonnegative integer $b$.
Then there exists a block-shuffle $w'$ of $w$ such that also $\pmax(w')\leq a$, but $w'$ has at most $a/2+2b+1$ blocks in each of the colors.
\end{lem}
\begin{proof}
Let us factorize $w$ as
$$w=w_1w_2\ldots w_n,$$
where $w_i$, for $i=1,2,\ldots,n$, are the blocks of $w$. Thus, odd-numbered blocks are colored with one color, while the even-numbered blocks are colored with the second color.
By considering swapping two consecutive blocks, we observe the following fact; the proof is a straightforward check.

\begin{clm}\label{cl:swap}
Suppose that for some $i$, $1\leq i<n$, we have that $\wsum(w_i)\geq 0$ and $\wsum(w_{i+1})\leq 0$. If $w'$ is obtained from $w$ by swapping blocks $w_i$ and $w_{i+1}$, 
then $w'$ is a block-shuffle of $w$ with $\pmax(w')\leq \pmax(w)$.
\end{clm}

Starting with the original word $w$, we apply the operation of Claim~\ref{cl:swap} exhaustively, up to the point when it cannot be applied anymore, or we obtain a word with exactly two blocks. 
Note that this procedure ends after a finite number of steps, as each swap strictly reduces the total number of blocks in the word 
(here we use the fact that we stop the procedure one a word with two blocks is obtained).
Let $w'$ be the word obtained at the end of this procedure. If $w'$ has one block of each color, then we are done. 
Otherwise, suppose $w'$ is such that the operation of Claim~\ref{cl:swap} cannot be applied to $w'$.
Then we have that $w'$ is a block-shuffle of $w$ and $\pmax(w')\leq \pmax(w)$.

Let us factorize $w'$ as
$$w'=w'_1w'_2\ldots w'_{n'},$$
where $w'_i$ are the blocks of $w'$. Let $I\subseteq \{1,2,\ldots,n'\}$ be the set of those positions $i$, for which $\wsum(w_i)\geq 0$.
Since the operation of Claim~\ref{cl:swap} is not applicable to $w'$, we infer that $I$ is a suffix of $\{1,2,\ldots,n'\}$, that is, there is a position $j$ such that $I=\{j,\ldots,n'\}$.
By the definition of $I$, we have $\wsum(w'_i)<0$ for all $i<j$.
Therefore, among blocks $w'_i$ for $i<j$ there can be at most $b$ blocks of each color; otherwise, restricting $w'$ (equivalently $w$) to letters of this color would yield a word with prefix minimum lower than $-b$.
Hence there can be at most $2b$ blocks before block $w_j$, and moreover we must have $\wsum(w'_1w'_2\ldots w'_{j-1})\geq -2b$.
On the other hand, for all $i>j$ we have that $\wsum(w_i)>0$, because otherwise the operation of Claim~\ref{cl:swap} would be applicable to blocks $w'_{i-1}$ and $w'_i$ (recall that $j<i$ implies that $i-1\in I$,
which means that $\wsum(w_{i-1})\geq 0$). Hence there cannot be more than $a+2b$ blocks after block $w_j$, because then we would have that $\pmax(w')>a$, 
contradicting $\pmax(w')\leq \pmax(w)\leq a$. We conclude that $w'$ has at most $a+4b+1$ blocks in total, so at most $a/2+2b+1$ blocks in each of the colors, as requested.
\end{proof}

\subsection{Factorizations: additional properties}\label{sec:factorizations-appendix}

In the following, we will use the fact that for any subset of nodes $U$ in a rooted forest, the numbers of factors in the maximal factorizations of $U$ and of the complement of $U$ are related to each other.
We begin with the case when the complement of $U$ is small.

\begin{lem}\label{lem:complement-bound}
Suppose $(X,U)$ is a partition of the node set of a rooted forest $F$.
Then the maximal factorization of $U$ has at most $|X|+1$ forest factors and at most $2|X|-1$ context factors.
\end{lem}
\begin{proof}
For a forest factor $A$ in factorization $\refine(U)$, let $p(A)$ be the parent of the roots of $A$, or $\bot$ if the roots of $A$ are root nodes. Observe that
whenever $p(A)$ is not equal to $\bot$, then it must be a node that belongs to $X$. Indeed, otherwise $A\cup \{p(A)\}$ would be a $U$-factor:
either a forest factor if all the children of $p(A)$ are in $A$, or a context factor if some do not belong to $A$. This would contradict the maximality of $A$.
Moreover, the function $p$ is injective, since if we had $p(A)=p(A')$ for different maximal $U$-factors $A,A'$ that are forest factors, then $A\cup A'$ would be also a $U$-factor, yielding that $A=A\cup A'=A'$ by the
maximality of $A$ and $A'$. We conclude that $p$ injectively maps the forest factors of $\refine(U)$ into $X\cup \{\bot\}$, hence there are at most $|X|+1$ forest factors in $\refine(U)$.

For a context factor $B$ in factorization $\refine(U)$, let $r(B)$ be the parent of the appendices of $B$. Clearly, since $r(B)\in B$, function $r$ is injective.
We prove that for a context factor $B$, either $r(B)$ has a child that belongs to $X$, or there are at least two different children of $r(B)$ that have nodes of $X$ as descendants.
Suppose the contrary. This implies that either all the descendants of $r(B)$ belong to $U$, 
or all the descendants of $r(B)$ that are contained in $X$ actually belong to the tree factor at the same child $u$ of $r(B)$, which moreover belongs to $U$.
In the first case we observe that by adding all the descendants of $r(B)$ to $B$ we obtain a tree factor that is a $U$-factor, which contradicts the maximality of $B$.
In the second case we observe that by adding to $B$ all the descendants of $r(B)$ apart from strict descendants of $u$ (in particular we add $u$) we obtain a context factor that is a $U$-factor, which 
again contradicts the maximality of $B$.

Therefore, $r$ injectively maps the context factors of $\refine(U)$ to the set consisting of parents of vertices of $X$ and lowest common ancestors of pairs of vertices of $X$.
It is well known that in a rooted forest, for any node subset $X$, the set of the lowest common ancestors of pairs of vertices from $X$ has size at most $|X|-1$.
Hence, $r$ injectively maps the context factors of $\refine(U)$ into a set of cardinality at most $2|X|-1$, thereby proving that the number of context factors in $\refine(U)$ is at most $2|X|-1$.
\end{proof}

Lemma~\ref{lem:complement-bound} can be conveniently lifted to the setting where $X$ can be large, but its maximal factorization has a small number of factors.

\begin{lem}[Lemma~\ref{lem:complement-full-bound-main}, restated]\label{lem:complement-full-bound}
Suppose $(U,W)$ is a partition of the node set of a rooted forest $F$, and let $k$ be the number of factors in the maximal factorization of $W$.
Then the maximal factorization of $U$ has at most $k+1$ forest factors and at most $2k-1$ context factors.
\end{lem}
\begin{proof}
Define a rooted forest $F'$ by identifying every maximal $W$-factor into a single vertex. 
More precisely, for every maximal $W$-factor $A$ that is a forest factor, replace it with a single node $x_A$. Make $x_A$ a child of the parent of the roots of $A$, or a root node if the roots of $A$ were root nodes.
Similarly, for every maximal $W$-factor $B$ that is a context factor, replace it with a single node $x_B$. Make $x_B$ a child of the parent of the root of $B$, or a root node if the root of $B$ was a root node.
Also, make every appendix of $B$ a child of $x_B$.
Let $X=\{x_A\colon A\in \refine(W)\}$, then $|X|=k$. 
It can be easily seen that every maximal $U$-factor in $F$ remains a maximal $U$-factor in $F'$.
Then the claim follows from Lemma~\ref{lem:complement-bound} applied to forest $F'$ and the partition $(X,U)$ of its node set.
\end{proof}

Finally, we observe that removing a small number of vertices from a set does not change the number of factors in its maximal factorization by much.

\begin{lem}\label{lem:max-fac-remove}
Suppose $F$ is a rooted forest and $U'\subseteq U$ are two node subsets such that $|U\setminus U'|\leq \ell$, for some nonnegative integer $\ell$.
Then
$$|\refine(U')|\leq 9|\refine(U)|+3\ell.$$
\end{lem}
\begin{proof}
Let $W=V(F)\setminus U$ and $W'=V(F)\setminus U'$ be the complements of $U$ and $U'$, respectively.
By Lemma~\ref{lem:complement-full-bound}, we have that
$$|\refine(W)|\leq 3|\refine(U)|.$$
Observe now that $W'=W\cup (U\setminus U')$, so there is a partition of $W'$ into $|\refine(W)|+|U\setminus U'|$ many $W'$-factors: one can take the maximal factorization of $W$ and add every
vertex of $U\setminus U'$ as a singleton factor. Consequently, the maximal factorization of $W'$ has at most this many factors, hence
$$|\refine(W')|\leq |\refine(W)|+|U\setminus U'|\leq |\refine(W)|+\ell.$$
Finally, $U'$ is the complement of $W'$, so using Lemma~\ref{lem:complement-full-bound} again we obtain that
$$|\refine(U')|\leq 3|\refine(W')|.$$
By combining the three inequalities above we are done.
\end{proof}

\subsection{Elimination forests}\label{sec:decompositions-appendix}

We begin with proving Lemma~\ref{lem:sep-forest-opt-main}, then we introduce reduced elimination forests and investigate their properties.

\begin{lem}[Lemma~\ref{lem:sep-forest-opt-main}, restated]\label{lem:sep-forest-opt}
For every graph $G$ there exists an elimination forest of $G$ whose width is equal to the treewidth of $G$.
\end{lem}
\begin{proof}
Let $t$ be an optimum-width tree decomposition of $G$.
Fix any linear order $\preceq$ on the vertices of $G$.
For every vertex $u$ of $G$, let $x_u$ be the node of $t$ whose margin contains $u$; since margins form a partition of the vertex set, such a node exists and is unique.
Define now a structure of a rooted forest $F$ on the vertex set of $G$ as follows:
\begin{itemize}
\item For any $u,v\in V(G)$, if $x_u$ is a strict ancestor of $x_v$ in $t$, then make $u$ an ancestor of $v$ in~$F$.
\item For any $u,v\in V(G)$, if $x_u=x_v$, then make $u$ an ancestor of $v$ in $F$ if $u\preceq v$, and make $v$ an ancestor of $u$ otherwise.
\end{itemize}
First observe that the forest $F$ defined above is an elimination forest of $G$. 
Indeed, for every vertex $u$, all the nodes of $t$ whose bags contain $u$ are descendants of $x_u$; this follows from the definition of the margin.
Hence, if $uv$ is an edge of $G$, then the node whose bag contains $u$ and $v$ must be both a descendant of $x_u$ and a descendant of $x_v$.
Consequently, $x_u$ and $x_v$ must be bound by the ancestor-descendant relation.

Finally, we verify that the width of the tree decomposition $t'$ induced by $F$ is no larger than the width of $t$.
To this end, we show that for every vertex $u$ of $G$, the bag of $u$ in $t'$ is a subset of the bag of $x_u$ in $t$.
Recall that the bag of $u$ in $t'$ consists of $u$ and all the ancestors of $u$ in $F$ that have a neighbor among the descendants of $u$ in $F$.
Clearly, $u$ itself belongs to the bag of $x_u$ in $t$.
Take then any vertex $v$ that is an ancestor of $u$ in $F$ that is adjacent to some $w$ that is a descendant of $u$ in $F$.
Since $v$ is an ancestor of $u$ in $F$ and $w$ is a descendant of $u$ in $F$, it follows that $x_v$ is an ancestor of $x_u$ in $t$ and $x_w$ is a descendant of $x_u$ in $t$.
The latter conclusion implies that all the nodes of $t$ whose bags contain $w$, are in fact descendants of $x_u$.
Observe that one of these bags must contain $v$ as well, as $vw$ is an edge of $G$.
Consequently, $v$ is contained both in the bag of some descendant of $x_u$, and in the bag of some ancestor of $x_u$, namely $x_v$.
This implies that $v$ is contained in the bag of $x_u$, as claimed.
\end{proof}

\paragraph*{Reduced elimination forests.} Intuitively, a reduced elimination forest is one that is minimal in terms of the depth of the nodes.

\begin{defi}
An elimination forest $F$ of a graph $G$ is {\em{reduced}} if for every vertex $u$ and every its child $v$ in $F$, $u$ has a neighbor among the descendants of $v$.
\end{defi}

The condition above was already considered in the context of treedepth~\cite{FominGP15} and trivially perfect graphs~\cite{DrangeFPV15}. 
We now show that in Lemma~\ref{lem:sep-forest-opt} one can require that the elimination forest is reduced.

\begin{lem}\label{lem:sep-forest-red-opt}
For every graph $G$ there exists a reduced elimination forest of $G$ whose width is equal to the treewidth of $G$.
\end{lem}
\begin{proof}
Lemma~\ref{lem:sep-forest-opt} asserts that there are some elimination forests of $G$ that have width equal to the treewidth of $G$.
Among these elimination forests, pick one that minimizes the sum of depth of all the vertices, and call it $F$.
We claim that $F$ is reduced.

Suppose, for the sake of contradiction, that some vertex $u$ has a child $v$ such that no descendant of $v$ is adjacent to $u$.
Modify $F$ by re-attaching $v$: make $v$ a child of the parent of $u$, instead of $v$, or make it into a root if $v$ has no parent (is a root).
Since we assumed that the descendants of $v$ are non-adjacent to $u$, it follows that the obtained forest $F'$ is still an elimination forest.
Moreover, during the modification only some vertices ceased to be the descendants of $u$, and otherwise the sets of ancestors and descendants of all the vertices stayed the same.
Consequently, in the construction of the induced tree decomposition from $F'$, every vertex will be assigned a bag that is a subset of the bag that was assigned to it when $F$ was considered.
This implies that the width of $F'$ is not larger than the width of $F$.
However, $F'$ has strictly smaller sum of depths of all the nodes than $F$. This contradicts the choice of $F$.
\end{proof}

We now derive a simple, yet useful property of a reduced elimination forest.

\begin{lem}\label{lem:connected}
Suppose $F$ is a reduced elimination forest of a graph $G$.
Then for every tree factor $A$ in $F$, the subgraph $G[A]$ is connected.
\end{lem}
\begin{proof}
For the sake of contradiction, suppose $A$ can be partitioned into nonempty subsets $X$ and $Y$ such that there is no edge between $X$ and $Y$ in $G$.
Since $A$ is a tree factor in $F$, there is at least one pair of vertices $(u,v)$ such that $u$ is the parent of $v$, while $u$ and $v$ belong to the opposite sides of the partition $(X,Y)$.
Choose $(u,v)$ so that $v$ is the deepest among pairs with this property, and assume w.l.o.g.\ that $u\in X$ and $v\in Y$.
By the choice of $(u,v)$, the tree factor at $v$ is entirely contained in $Y$. 
Hence no descendant of $v$ is adjacent $u$, due to $u\in X$.
This is a contradiction with $F$ being reduced.
\end{proof}

Finally, we derive two additional technical lemmas about reduced tree decompositions, which will be needed to achieve property~\ref{it:few-context-factors} of the Dealternation Lemma.

\begin{lem}\label{lem:context-bound}
Suppose $F$ is a reduced elimination forest of a graph $G$; let $\ell$ be the width of $F$.
Suppose further that $X,A_1,A_2,\ldots,A_p$ is a partition of the vertex set of $G$ such that there is no edge between $A_i$ and $A_j$ for $i\neq j$.
Then any maximal $(V(G)\setminus X)$-factor that is a context factor, intersects at most $\ell+1$ among sets $A_1,A_2,\ldots,A_p$.
\end{lem}
\begin{proof}
Let $t$ be the tree decomposition induced by $F$.
Fix any maximal $(V(G)\setminus X)$-factor $B$ that is a context factor, and assume $B\cap A_i$ is nonempty for some $i$.

Let $B'\supseteq B$ be the tree factor whose root is the root of $B$.
Since $B$ is a maximal $(V(G)\setminus X)$-factor that is a context factor, we infer that $B'$ must contain at least one vertex of $X$, 
because otherwise $B'$ would be a $(V(G)\setminus X)$-factor that would be a strict superset of $B$.
By Lemma~\ref{lem:connected}, $G[B']$ is connected.
Observe that $X\cap B'\subseteq B'\setminus B$, and hence sets $X\cap B'$ and $A_i\cap B$ are disjoint.
Let $P$ be a shortest path between $X\cap B'$ and $A_i\cap B$ in $G[B']$.
Denote the endpoints of $P$ by $u$ and $v$, where $u\in A_i\cap B$ and $v\in X\cap B'$.
As $P$ was chosen to be the shortest, no vertex of $P$ apart from $v$ belongs to $X$.
Since all the neighbors of vertices of $A_i$ lie in $A_i\cup X$, and $u$ belongs to $A_i$, we infer that all the vertices on $P$ apart from $v$ belong to $A_i$.

Since $P$ is connected, the set of those vertices whose bags in $t$ contain any vertex of $P$, induces a connected subtree of $F$.
This subtree contains both a vertex in $B$, namely $u$, and a vertex in $B'\setminus B$, namely $v$, and hence it contains the whole path in $F$ between these vertices.
In particular the parent of the appendices of $B$ is included in this subtree; denote it by $w$.
Summarizing, there is a vertex $a$ on $P$ that is included in the bag of $w$.
Observe that $a$ cannot be equal to $v$. This is because $v$ belongs to the forest factor $B'\setminus B$, so all the nodes whose bags contain $v$ also belong to this forest factor.
Consequently, $a$ is a vertex on $P$ that is different than $v$, so $a\in A_i$.

Since this reasoning can be performed for each $i$ such that $B\cap A_i$ is nonempty, for each such index $i$ we obtain a different vertex $a$ that needs to be included in the bag of $w$.
The size of the bag of $w$ is, however, bounded by $\ell+1$, so the same bound holds also for the number of indices $i$ as above.
\end{proof}

\begin{lem}\label{lem:children-bound}
Suppose $t$ is a tree decomposition of width $k$ of a graph $G$ and $F$ is a reduced elimination forest of $G$ of width at most $k$, such that $t$ and $F$ satisfy condition~\dlref{it:few-factors}
of the Dealternation Lemma (Lemma~\ref{lem:dealternation}) for some function $f(k)\in \Oh(k^3)$.
Then for every node $x$ of $t$, there are at most $g(k)$ children of $x$ in the set
$$\{ y \colon y\textrm{ is a node of }t\textrm{ with at least one context factor in }\refine_F(\comp_t(y))\},$$
where $g(k)\in \Oh(k^4)$ is a function depending on $f(k)$ only.
\end{lem}
\begin{proof}
Fix any node $x$ of $t$, and let $y_1,y_2,\ldots,y_p$ be its children in $t$. Denote 
$$A_i=\comp_t(y_i)\ \textrm{ for all }\ i=1,2,\ldots,p,\quad\textrm{ and }\quad X=V(G)\setminus \bigcup_{i=1}^p A_i.$$
Since $t$ is a tree decomposition of $G$, it follows that there is no edge between $A_i$ and $A_j$ for any $i\neq j$, and hence the tuple $(X,A_1,A_2,\ldots,A_p)$ satisfies the prerequisites
of Lemma~\ref{lem:context-bound}. 
Recall that $F$ is reduced and has width at most $k$, so by Lemma~\ref{lem:context-bound} we conclude that for any context factor $B$ from the maximal factorization of $\refine_{F}(V(G)\setminus X)$, 
at most $k+1$ among sets $A_1,A_2,\ldots,A_p$ intersect $B$.
Note that $V(G)\setminus X=\comp_t(x)\setminus \mar_t(x)$, so $V(G)\setminus X$ can be obtained from $\comp_t(x)$ by removing at most $k+1$ vertices.
The maximal factorization of $\comp_t(x)$ in $F$ has at most $f(k)$ factors, 
so by Lemma~\ref{lem:max-fac-remove} we have that the maximal factorization of $V(G)\setminus X$ in $F$ has at most $9\cdot f(k)+3(k+1)$ factors.
Consequently, if we take 
$$g(k)=(9\cdot f(k)+3(k+1))\cdot (k+1),$$
then at most $g(k)$ among sets $A_1,A_2,\ldots,A_p$ can intersect any context factor in the maximal factorization of $V(G)\setminus X$. 
We claim that all the other sets $A_i$ have only forest factors in their maximal factorizations, which will conclude the proof.

Take any such $A_i$, that is, $A_i$ intersects only forest factors of the maximal factorization of $V(G)\setminus X$.
Let $B$ be any tree factor in $F$ that is contained in $V(G)\setminus X$.
Since $F$ is reduced, by Lemma~\ref{lem:connected} we have that $G[B]$ is connected.
There are no edges between $A_i$ and $A_j$ for any $j\neq i$, so we conclude that $B$ is either entirely contained or entirely disjoint with $A_i$. 
Since $A_i$ is disjoint with all the context factors of $\refine_F(V(G)\setminus X)$, it follows that the set $A_i$ is closed under taking descendants in $F$.
In particular, this implies that the maximal factorization of $A_i$ contains no context factors, as promised.
\end{proof}

\subsection{From Local to Global Dealternation Lemma}\label{sec:dealternation-apendix}

In this section we give a proof of the Dealternation Lemma assuming its local counterpart, which will be formulated in a moment.
First, for convenience we introduce the appropriate notion of alternation for tree decompositions.

\begin{defi}
Suppose $t$ is a tree decompositions of a graph $G$, and $F$ is an elimination forest of~$G$.
The {\em{$t$-alternation of $F$}} is defined as the maximum among the nodes $x$ of $t$, of the number of maximal $\comp_{t}(x)$-factors in $F$.
In other words, the $t$-alternation of $F$ is equal to:
$$\max_{x\in V(t)} |\refine_F(\comp_t(x))|.$$
\end{defi}

Thus, to prove the Dealternation Lemma it suffices to show that there always exists an optimum-width elimination forest $F$ of $G$,
such that the $t$-alternation of $F$ is bounded by a quadratic function of the width of $t$ (that is, condition~\dlref{it:few-factors} holds), and such that $F$ also
satisfies condition~\dlref{it:few-context-factors}.

The idea for the proof is as follows. 
We take any reduced elimination forest $F$ of $G$ of optimum width, and iteratively ``correct'' $F$ so that its $t$-alternation becomes bounded.
To achieve this, we examine each node $x$ of $t$ and correct $F$ so that the number of $\comp_{t}(x)$-factors in $F$ is bounded by $f(k)$, for some quadratic function $f$.
For this, we devise a local correction procedure, which we call the Local Dealternation Lemma; this procedure is applied iteratively to all the nodes of $t$.

\newcommand{\ldl}{LD}
\newcommand{\ldlref}[1]{\ref{#1}}

\begin{lem}[Local Dealternation Lemma]
There exists a function $f(k)\in \Oh(k^3)$ such that the following holds.
Suppose $G$ is a graph of treewidth at most $k$, and $F$ is a reduced elimination forest of $G$ of optimum width.
Suppose further that $(U,X,W)$ is a partition of the vertex set of $G$ such that $|X|\leq k+1$ and there is no edge between $U$ and $W$.
Then there is a reduced elimination forest $F'$ of $G$ of optimum width with the following conditions satisfied:
\begin{enumerate}[(\ldl 1)]
\item\label{c:fewU} There are at most $f(k)$ maximal $U$-factors in $F'$.
\item\label{c:Ucont} For every $U'\subseteq U$, every $U'$-factor in $F$ is also a $U'$-factor in $F'$.
\item\label{c:Wcont} For every $W'\subseteq W$, every $W'$-factor in $F$ is also a $W'$-factor in $F'$.
\end{enumerate}
\end{lem}

We remark that a statement formulating the essence of the Local Dealternation Lemma can be found in the work of Courcelle and Lagergren~\cite[Theorem 6.3]{CourcelleL96}.

When applying the Local Dealternation Lemma to each component of $t$, we need to be careful, as we have to make sure that one application that corrects $\comp_t(x)$-factors for some $x$, does not
increase the number of $\comp_t(y)$-factors for nodes $y$ that were corrected before. 
To achieve this, we shall apply the Local Dealternation Lemma in a bottom-up order on the nodes of $x$.
At the end, this ensures that property~\dlref{it:few-factors} is satisfied.
For property~\dlref{it:few-context-factors}, we guarantee that all the intermediate, as well as the final elimination forest is reduced, and we make use of Lemma~\ref{lem:context-bound}.
We now proceed to a formal reasoning, supposing that the Local Dealternation Lemma holds.

\begin{proof}[Proof of Dealternation Lemma, using Local Dealternation Lemma]
Since we know that $t$ has width at most~$k$, we have that all adhesions in $t$ have sizes not larger than $k+1$.
Let $F_0$ be any reduced elimination forest of $G$ of optimum width, which exists by Lemma~\ref{lem:sep-forest-red-opt}.
Clearly, as $t$ has width~$k$, the width of $F_0$ is at most $k$.

Let $\preceq$ be an arbitrary linear order on the node set of $t$ such that whenever a node $x$ is a strict descendant of a node $y$, then $x$ comes before $y$ in $\preceq$.
Let 
$$V(t)=\{x_1\prec x_2\prec\ldots\prec x_m\},$$
where $m=|V(t)|$. We process the nodes of $t$ in the order $\preceq$, inductively computing reduced elimination forests $F_1,\ldots,F_m$, starting with $F_0$.
We keep the following invariant for every $i=0,1,\ldots,m$: in decomposition $F_i$, the number of $\comp_t(x_j)$-factors is at most $f(k)$ for every $j\leq i$, where $f$ is the function
given by the Local Dealternation Lemma.
Thus, the invariant is satisfied vacuously for $i=0$. 
Observe that the reduced elimination forest $F_m$ obtained at the end of the construction has $t$-alternation bounded by $f(k)$.

For $i\geq 1$, construct decomposition $F_i$ by applying the Local Dealternation Lemma to the elimination forest $F_{i-1}$ and partition 
$$(U,X,W):=(\comp_t(x_i),\adh_t(x_i),V(G)\setminus (\comp_t(x_i)\cup \adh_t(x_i)))$$
of the vertex set of $G$; the fact that this partition satisfies the prerequisites of the Local Dealternation Lemma follows from the properties of the tree decomposition $t$. 
Clearly, by condition~\ldlref{c:fewU}, the number of maximal $\comp_t(x_i)$-factors in $F_i$ is at most $f(k)$.
It remains to prove the same conclusion for maximal $\comp_t(x_j)$-factors, for every $j<i$. Since $x_j\prec x_i$, we have that $x_j$ is not an ancestor of $x_i$. 

If $x_j$ is a descendant of $x_i$, then $\comp_t(x_j)\subseteq \comp_t(x_i)=U$.
By condition~\ldlref{c:Ucont}, every $\comp_t(x_j)$-factor in $F_{i-1}$ is also an $\comp_t(x_j)$-factor in $F_i$, hence the number of maximal $\comp_t(x_j)$-factors in $F_i$ cannot be larger
than the number of maximal $\comp_t(x_j)$-factors in $F_{i-1}$, which is at most $f(k)$ by induction.

If $x_j$ is not a descendant of $x_i$, then since it is neither an ancestor, we obtain that $\comp_t(x_j)\subseteq V(G)\setminus (\comp_t(x_i)\cup \adh_t(x_i))=W$.
Again, by condition~\ldlref{c:Wcont}, every $\comp_t(x_j)$-factor in $F_{i-1}$ is also an $\comp_t(x_j)$-factor in $F_i$, hence again the number of maximal $\comp_t(x_j)$-factors in $F_i$ cannot be larger than
$f(k)$ by induction.

Thus, we have found an elimination forest $F=F_m$ of $G$ such that: (i) $F$ is reduced and has optimum width, and (ii) the $t$-alternation of $F$ is bounded by $f(k)$.
Hence, $F$ satisfies property~\dlref{it:few-factors}.
Finally, note that property~\dlref{it:few-context-factors} is also satisfied for $F$ due to Lemma~\ref{lem:children-bound}, because $F$ is reduced.
\end{proof}

\subsection{Proof of the Local Dealternation Lemma}\label{sec:local-dealternation-apendix}

We are left with proving the Local Dealternation Lemma. 
Let $F$ be the given reduced elimination forest of $G$ of optimum width.
Also, let $s$ be the tree decomposition induced by $F$. Recall that the forest underlying $s$ is equal to $F$, while the bags are constructed as described in Section~\ref{sec:dealternation}.
Finally, let $\ell\leq k$ be the width of $s$, which is equal to the treewidth of $G$. 

To ease the description, we color the vertices of $G$ as follows:
vertices of $U$ are \textcolor{red}{red}, and the vertices of $W$ are \textcolor{blue}{blue}. The vertices of $X$ do not receive any color.
When we say that some set is {\em{monochromatic}}, we mean that all its members are red or all its members are blue. 
In particular, a monochromatic set has no elements of $X$. Similarly, when we say that two vertices have the same color, or are of different colors,
we implicitly state that both of them are assigned some color, so they belong to $U\cup W$.

\newcommand{\invr}{{I}}
\newcommand{\invref}[1]{\ref{#1}}

The idea is to modify the forest $F$ by performing local ``surgery'' on its shape, so that at the end it satisfies condition~\ldlref{c:fewU}.
During the modification we will make sure that the final decomposition will be reduced and will satisfy conditions~\ldlref{c:Ucont} and~\ldlref{c:Wcont}.
In order not to obfuscate the description, we do not verify conditions~\ldlref{c:Ucont} and~\ldlref{c:Wcont} directly, 
as their satisfaction follows immediately from the nature of the modification performed.
More precisely, the modification will satisfy the following invariants:
\begin{enumerate}[(\invr 1)]
\item\label{i:parent-child} Whenever $u$ is the parent of $v$, and $u$ and $v$ are of the same color, then $u$ remains the parent of $v$ after the modification.
\item\label{i:siblings} Whenever $u$ and $v$ are siblings, and the tree factors at $u$ and $v$ are monochromatic and of the same color, then $u$ and $v$ remain siblings after the modification.
\item\label{i:leaf} Whenever $u\in U\cup W$ is a leaf, it remains a leaf after the modification.
\end{enumerate}
\noindent It is easy to see that the satisfaction of these invariants ensures that conditions~\ldlref{c:Ucont} and~\ldlref{c:Wcont} are preserved.
We leave the verification of the invariants throughout the description to the reader.
Finally, the fact that the output elimination forest is reduced will be checked explicitly.

The main idea is to examine each maximal $(U\cup W)$-factor in $F$, and reorganize it so that it can be partitioned into a bounded number of $U$- and $W$-factors.
The fact that $F$ is reduced implies that no reorganization is needed for forest factors: from Lemma~\ref{lem:connected} it follows that every forest factor of $\refine_F(U\cup W)$
can be partitioned into one $U$-factor and one $W$-factor. For context factors of $\refine_F(U\cup W)$, some rearrangement is, however, necessary.
For this, we will use the tools developed in Section~\ref{sec:words}.

We start by observing the following property that is implied by the fact that $F$ is reduced.

\begin{clm}\label{cl:cleaning-result}
Suppose $u$ is a vertex such that the tree factor at $u$ in $F$ is entirely contained in $U\cup W$.
Then this tree factor is monochromatic.
Moreover, if $u$ has a parent $v$, then it cannot happen that $u$ and $v$ have different colors.
\end{clm}
\begin{proof}
Let $A$ be the tree factor at $u$.
Since there is no edge between $U$ and $W$, there is also no edge between $U\cap A$ and $W\cap A$.
However, $G[A]$ is connected by Lemma~\ref{lem:connected}. Hence either $U\cap A$ or $W\cap A$ is empty, which establishes the first claim.
For the second claim, observe that otherwise the pair $(v,u)$ would contradict the fact that $F$ is reduced.
\cqed\end{proof}

We now examine $\refine_{F}(U\cup W)$, the maximal factorization of $U\cup W$ in $F$. Recall that this partition of $U\cup W$ consists of all maximal $(U\cup W)$-factors in $F$.
Since $|X|\leq k+1$, by Lemma~\ref{lem:complement-bound} we obtain the following.
\begin{clm}\label{cl:UW-small-number}
Factorization $\refine_{F}(U\cup W)$ has at most $k+2$ forest factors and at most $2k+1$ context factors.
\end{clm}

We now observe that Claim~\ref{cl:cleaning-result} already implies that the forest factors of $\refine_{F}(U\cup W)$ are as required:

\begin{clm}\label{cl:forest-fewU}
Let $A$ be a maximal $(U\cup W)$-factor in $F$ that is a forest factor. Then each of $A\cap U$ and $A\cap W$ is either empty or a forest factor in $F$.
\end{clm}
\begin{proof}
By Claim~\ref{cl:cleaning-result}, every tree factor contained in $A$ is monochromatic. 
Consequently, $A\cap U$ consists of all the tree factors at the red roots of $A$, thus it is either empty or a forest factor.
An analogous arguments applies to $A\cap W$.
\cqed\end{proof}


The context factors of $\refine_{F}(U\cup W)$ may need reorganization. Fix some context factor $B$ from $\refine_{F}(U\cup W)$. 
We now analyze the structure of $B$. The path from the root of $B$ to the parent of the appendices of $B$ shall be called the {\em{spine}} of the context factor $B$; we denote the spine by $S$. For a vertex $v\in S$, let $R_v$ denote the set of those strict descendants of $v$ which belong to $B$, and for which $v$ is their lowest ancestor on the spine. 
Note that $R_v$ may be empty, if no such descendant exists, and otherwise it is a forest factor with roots being those children of $v$ that are in $B$ but are not on $S$. Let us observe the following.


\begin{clm}\label{cl:Rv-mono}
For each vertex $v\in S$, every vertex of $R_v$ has the same color as $v$.
\end{clm}
\begin{proof}
Follows immediately from Claim~\ref{cl:cleaning-result}.
\cqed\end{proof}

For $v\in S$, let $C_v=\{v\}\cup R_v$. Observe that each such set $C_v$, for $v\in S$, is a context factor in $F$, which is moreover monochromatic by Claim~\ref{cl:Rv-mono}.
Thus, each $C_v$ is a $U$- or $W$-factor, depending on the color of $v$.

A vertex $v\in S$ shall be called {\em{important}} if either $v$ is the deepest vertex on $S$ (i.e., the parent of the appendices of $B$), or $\adh_s(v)\setminus \adh_s(v')$ contains a vertex of $X$, where $v'$ is the child of $v$ on $S$. We note that there are not so many important vertices.

\begin{clm}\label{cl:important-bound}
 There are at most $\ell+1$ important vertices on $S$.
\end{clm}
\begin{proof} 
For each important vertex $v\in S$ that is not the deepest vertex on $S$, select any vertex $x_v$ that belongs both to $X$ and to $\adh_s(v)\setminus \adh_s(v')$, where $v'$ is the child of $v$ on $S$. Observe that since $B\cap X=\emptyset$, it follows that $x_v$ is an ancestor of all the vertices of $S$. Hence $x_v\in \adh_s(r)$, where $r$ is the root of $B$. As $|\adh_s(r)|\leq \ell$ and vertices $x_v$ are pairwise different for different vertices $v$, it follows that the number of important vertices on $S$ is at most $\ell+1$ (where the $+1$ summand is contributed by the deepest vertex of $S$). 
\cqed\end{proof}

\newcommand{\Ff}{\mathcal{F}}

We now consider a factorization $\Ff_B$ of $B$ into context factors defined as follows:
\begin{itemize}
 \item For each important vertex $v$ on $S$, put $C_v$ into $\Ff$ as a separate context factor. These context factors shall be called {\em{important}}.
 \item For each maximal subpath $S'$ of $S$ that does not contain any important vertices, put the context factor $\bigcup_{v\in S'} C_v$ into $\Ff$. These context factors shall be called {\em{regular}}.
\end{itemize}
By Claim~\ref{cl:important-bound}, $\Ff_B$ consists of at most $\ell+1$ important factors and at most $\ell+1$ regular factors. The important factors of $\Ff_B$ are monochromatic by Claim~\ref{cl:Rv-mono}, but the same cannot be said about the regular ones. Therefore, let us fix a regular factor $B'\in \Ff_B$. That is, $B'=\bigcup_{v\in S'} C_v$ for some maximal subpath $S'$ of $S$ that does not contain any important vertices. The path $S'$ shall be called the spine of $B'$.

Let us enumerate the vertices of $S'$ as $v_1,\ldots,v_m$, where $v_i$ is an ancestor of $v_j$ for $i\leq j$. Further, let $v_{m+1}$ be the child of $v_m$ on $S$ (which exists due to considering the deepest vertex of $S$ to be important). For brevity, we will write $R_i=R_{v_i}$ and $C_i=C_{v_i}$.


For each $i\in \{1,2,\ldots,m\}$, define
$$Q_i=\adh_s(v_i)\setminus \adh_s(v_{i+1}).$$
Note that by the way $s$ is constructed from $F$ (see Section~\ref{sec:dealternation}), $Q_i$ comprises all strict ancestors of $v_i$ that do have a neighbor in $C_i$, but do not have any neighbors among the descendants of $v_{i+1}$. This implies the following.

\begin{clm}\label{cl:Q-mono}
 For each $i\in \{1,2,\ldots,m\}$, if $v_i\in U$ then $Q_i\subseteq U$, and if $v_i\in W$ then $Q_i\subseteq W$.
\end{clm}
\begin{proof}
By symmetry, suppose $v_i\in U$. Consider any $x\in Q_i$ and let $w$ be any neighbor of $x$ in $C_i$. By Claim~\ref{cl:Rv-mono}, we have $w\in U$. Since $v_i$ is not important, $x\notin X$. Therefore we must have $x\in U$, for otherwise $wx$ would be an edge with one endpoint in $U$ and the second in $W$.
\cqed\end{proof}

\begin{figure}[htbp!]
        \centering
                \def\svgwidth{\textwidth}
                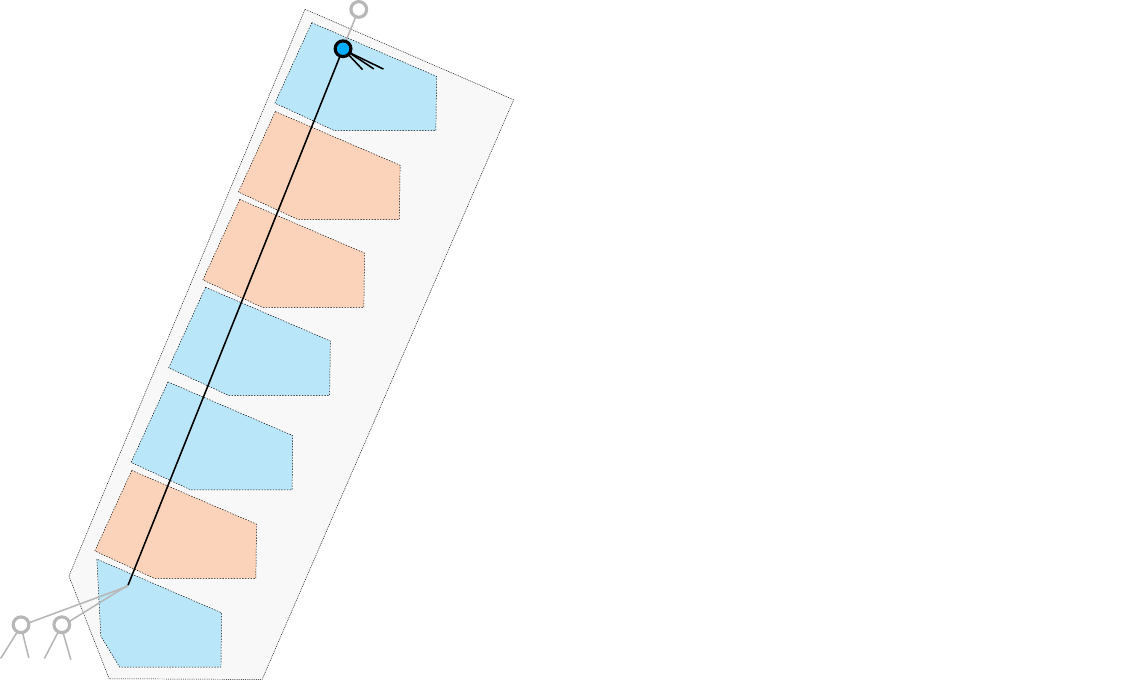
\caption{Reorganization of an example regular factor $B'$ with $7$ vertices on the spine.
The context before the reorganization is on the left panel, after is on the right. 
The applied permutation is $\pi=(1,4,5,2,3,6)$, and it leaves only two monochromatic blocks in $h'$.
Note that the last context $C_7$ does not participate in the reorganization and stays on its place.}\label{fig:context-reorganization}
\end{figure}

For $i=1,2,\ldots,m$, let $x_i$ be the word over the alphabet $\Sigma=\{+,-\}$ defined as follows:
$$x_i=+(-)^{|Q_i|}.$$
That is, we first put one $+$, and then repeat $-$ exactly $|Q_i|$ times.
Color $x_i$ with the same color as~$v_i$, and define a bichromatic word $h$ as follows:
$$h=x_1x_2\ldots x_m.$$
The idea is apply the block-shuffle given by Lemma~\ref{lem:reorder-two} to $h$; this block-shuffle will naturally induce a reorganization of $B'$ within $F$, as depicted on Figure~\ref{fig:context-reorganization}.
Thus, the number of monochromatic blocks will be reduced, while the additional properties asserted by Lemma~\ref{lem:reorder-two} will ensure that the width of the decomposition does not increase.

We proceed to the details, but first we need to examine the parameters of $h$ needed to apply Lemma~\ref{lem:reorder-two}.

\begin{clm}\label{cl:telescope}
For each $i\in \{1,2,\ldots,m\}$, we have $\wsum(x_i)=|\adh_{s}(v_{i+1})|-|\adh_{s}(v_i)|$.
\end{clm}
\begin{proof}
Observe that $\adh_{s}(v_{i+1})\subseteq \adh_s(v_i)\cup \{v_i\}$ and $\adh_s(v_{i+1})\setminus \adh_s(v_i)=\{v_i\}$, because $F$ is reduced. Therefore
\begin{eqnarray*}
\wsum(x_i) & = & 1-|Q_i|\\
           & = & 1-|\adh_s(v_i)\setminus \adh_{s}(v_{i+1})|\\
           & = & 1-(|\adh_{s}(v_{i})|-|\adh_{s}(v_{i+1})|+1)\\
           & = & |\adh_{s}(v_{i+1})|-|\adh_{s}(v_i)|, 
\end{eqnarray*}
as claimed.
\cqed \end{proof}

\begin{clm}\label{cl:reshuffle-prerequisites}
If $h_{\textrm{red}}$ and $h_{\textrm{blue}}$ are formed by restricting $h$ to red and blue letters, then
$$\pmax(h)\leq \ell+1-|\adh_{s}(v_1)|\quad\textrm{and}\quad \pmin(h_{\textrm{red}})\geq -\ell\quad \textrm{and} \quad \pmin(h_{\textrm{blue}})\geq -\ell.$$
\end{clm}
\begin{proof}
Observe since each subword $x_i$ contains only one $+$, we have that 
$$\pmax(h)\leq 1+\max_{i=0,1,2,\ldots,m-1}\wsum(x_1x_2\ldots x_i).$$
On the other hand, by Claim~\ref{cl:telescope}, for every $i=0,1,\ldots,m$ we have that
$$\wsum(x_1x_2\ldots x_i)=|\adh_{s}(v_{i+1})|-|\adh_{s}(v_1)|\leq \ell-|\adh_{s}(v_1)|.$$
The first claimed inequality follows.

Consider now the word $h_{\textrm{red}}$, which is defined as
$$h_{\textrm{red}}=x_1'x_2'\ldots x_m',$$
where $x_i'=x_i$ if $v_i$ is red, and $x'_i=\varepsilon$ if $v_i$ is blue. Clearly, we have
$$\pmin(h_{\textrm{red}})=\min_{i=0,1,\ldots,m}\wsum(x'_1x'_2\ldots x'_i).$$
On the other hand, since word $x'_i$ is nonempty exactly when $C_i$ is red, similarly as in Claim~\ref{cl:telescope} we obtain that
$$\wsum(x'_i)=|\adh_{s}(v_{i+1})\cap U|-|\adh_{s}(v_i)\cap U|.$$
Consequently, we have
$$\wsum(x'_1x'_2\ldots x'_i)=|\adh_{s}(v_{i+1})\cap U|-|\adh_{s}(v_{i})\cap U|\geq 0-\ell=-\ell,$$
which implies the second claimed inequality. The proof for the third one is analogous.
\cqed\end{proof}

Thus, we can apply Lemma~\ref{lem:reorder-two} to the word $h$, obtaining a word $h'$ with the following properties.
\begin{itemize}
\item Word $h'$ is a block-shuffle of $h$, in particular every subword $x_i$ remains contiguous in $h'$.
\item $\pmax(h')\leq \pmax(h)$.
\item The numbers of red and blue blocks in $h'$ are not larger than $(5\ell+3)/2$.
\end{itemize}
Now, based on $h'$, we construct the modified context factor in a natural manner.
Let $\pi\colon \{1,\ldots,m\}\to \{1,\ldots,m\}$ be a permutation such that 
$$h'=x_{\pi(1)}x_{\pi(2)}\ldots x_{\pi(m)}.$$
Then permute the context factors $\{C_i\colon i\in \{1,2,\ldots m\}\}$ according to $\pi$; see Figure~\ref{fig:context-reorganization} for an illustration.
\begin{itemize}
\item Make $v_{\pi(1)}$ into a child of the node that was the parent of $v_1$ in $s$; in case $v_1$ was a root node, $v_{\pi(1)}$ becomes a root node.
\item For each $i=2,3,\ldots,m$, make $v_{\pi(i+1)}$ a child of $v_{\pi(i)}$.
\item Make $v_{m+1}$ into a child of $v_{\pi(m)}$.
\end{itemize}
Since there is no edge between red and blue vertices in $G$, and $h'$ is a block-shuffle of $h$, this reorganization seems not to spoil the basic assumption that we are working with an elimination forest.
We now verify this formally.

Apply the reorganization defined above to every regular factor $B'$ belonging to the factorization $\Ff_B$, for every context factor $B\in \refine_{F}(U\cup W)$. Let $F'$ be the obtained rooted forest. We now verify the properties of $F'$. For convenience, denote 
$$\Ff=\bigcup_{B\in \refine_F(U\cup W)} \left(\textrm{the set of regular factors of }\Ff_B\right),$$
thus $\Ff$ comprises all the factors $B'$ to which the reorganization is applied.

\begin{clm}\label{cl:reorganize-correct}
$F'$ is an elimination forest of $G$.
\end{clm}
\begin{proof}
Take any edge $uv\in E(G)$. Since $F$ is an elimination forest of $G$, we have that $u$ and $v$ are bound by the ancestor-descendant relation in $F$; say $u$ is an ancestor of $v$.
If $u\notin \bigcup \Ff$ or $v\notin \bigcup \Ff$, then $u$ remains an ancestor of $v$ in $F'$, because the modification yielding $F'$ is performed in each factor of $\Ff$ separately, while the vertices outside of $\bigcup \Ff$ stay intact. 
Similarly, if $u\in B'_u\in \Ff$ and $v\in B'_v\in \Ff$ where $B'_u\neq B'_v$, then the relative positions of $B'_u$ and $B'_v$ do not change during the reordering, and $u$ remains an ancestor of $v$ in $F'$.

We are left with the case when $u$ and $v$ belong to the same factor $B'\in \Ff$. Note that in particular $u,v\notin X$.
Since $uv$ is an edge, it cannot be that $u\in U$ and $v\in W$ or vice versa. Assume then, without loss of generality, that $u,v\in U$, that is, both $u$ and $v$ are red. Let $u'$ and $v'$ be vertices on the spine of $B'$ such that $u\in C_{u'}$ and $v\in C_{v'}$. Note that $u'$ and $v'$ are both red. Further, since $u$ is an ancestor of $v$ in $F$, we either have $u'=v'$, or $u'$ is a strict ancestor of $v'$ in $F$ and $u=u'$. In the former case, the ancestor-descendant relation within $C_{u'}=C_{v'}$ is left intact by the reorganization, so $u$ remains an ancestor of $v$ in $F'$. In the latter case, since the reorganization within $B'$ is performed by a block-shuffle, the relative order of $u'=u$ and $v'$ on the spine does not change, so again $u$ remains an ancestor of $v$ in~$F'$.
\cqed\end{proof}


\begin{clm}\label{cl:reorganize-reduced}
$F'$ is reduced.
\end{clm}
\begin{proof}
Take any vertex $u$.
Observe that if $u$ does not lie on the spine of any factor of $\Ff$, then $u$ has exactly the same descendants in $F$ and in $F'$, which are moreover partitioned in the same manner
among the tree factors at the children of $u$.
Hence, it remains to check what happens if $u$ lies on the spine of a factor $B'\in \Ff$.

Adopt the notation from the description of the reorganization of the factor $B'$, and suppose $u=v_i$ for some $i\in \{1,2,\ldots,m\}$.
W.l.o.g.\ suppose $u$ is red.
Every tree factor contained in $R_i$ stays intact in $F'$ and remains attached below $u$, so $u$ has still a neighbor in each of these tree factors.
Therefore, the only tree factor at a child of $u$ in $F'$ that remains to be checked is the tree factor at the child
of $u$ on the spine of $B'$. Since the reorganization was obtained by a block-shuffle, the relative positions of red vertices along the spine remain the same in $F'$ as they were in $F$.
Hence, this tree factor is obtained from the tree factor at $v_{i+1}$ in $F$ by adding and/or removing some blue vertices.
Since $F$ was reduced, $u$ has a neighbor $w$ in the tree factor rooted at $v_{i+1}$ in $F$. 
The neighbor $w$ in particular cannot be blue, because $u$ is red.
We infer that $w$ remains in the tree factor at the child of $v_i$ on the spine in $F'$, which concludes the proof.
\cqed\end{proof}


\begin{clm}\label{cl:reorganize-width}
The width of $F'$ is not larger than the width of $F$.
\end{clm}
\begin{proof}
Let $s'$ be the tree decomposition induced by $F'$.
Take any vertex $v$ of $G$. If $v\notin \bigcup \Ff$, then $v$ has exactly the same ancestors and descendants in $F$ as in $F'$,
hence it is assigned exactly the same bag in the induced decompositions $s$ and $s'$. Therefore, from now on assume that $v\in B'$ for some $B'\in \Ff$. In particular $v\notin X$, hence assume without loss of generality that $v\in U$, i.e., $v$ is red.

Adopt the notation from the description of the reorganization of the factor $B'$.
Suppose first that $v$ does not lie on the spine of $B'$. In this case, the set of descendants of $v$ does not change during the reorganization, however $v$ can get new ancestors on the spine of $B'$.
Observe, nevertheless, that all these new ancestors will be blue, because the reorganization applied to the context factor $B'$ does not change the relative order of red vertices on the spine.
As all descendants of $v$ are red (by Claim~\ref{cl:Rv-mono}), we infer that none of the new ancestors of $v$ is included in the bag of $v$ in $s'$. Consequently, the bag of $v$ in $s'$ is a subset of the bag of $v$ in $s$.

Finally, we are left with the case when $v$ belongs to the spine of $B'$,  say $v=v_i$ for some $i\in \{1,2,\ldots,m\}$.
First,  we observe that for every $j\in \{1,2,\ldots,m\}$ it holds that
$$
\adh_{s}(v_j)\setminus \adh_{s}(v_{j+1}) = \adh_{s'}(v_j)\setminus \adh_{s'}(v_{\pi(\pi^{-1}(j)+1)}).
$$
This is because $\adh_{s}(v_j)\setminus \adh_{s}(v_{j+1})$ consists of strict ancestors of $v_j$ with the same color as $v_j$, which in particular do not belong to $X$ (Claim~\ref{cl:Q-mono}), and $B'$ is reorganized through a block shuffle.
Since $F$ and $F'$ are reduced (Claim~\ref{cl:reorganize-reduced}), we also have
$$
\adh_{s}(v_{j+1})\setminus \adh_{s}(v_j) = \adh_{s'}(v_{\pi(\pi^{-1}(j)+1)})\setminus \adh_{s'}(v_j)=\{v_j\}.
$$
By Claim~\ref{cl:telescope}, this implies that
\begin{equation*}
\wsum(x_j)=|\adh_{s}(v_{j+1})|-|\adh_{s}(v_{j})| = |\adh_{s'}(v_{\pi(\pi^{-1}(j)+1)})|-|\adh_{s'}(v_{j})|.
\end{equation*}
Therefore, we have
\begin{equation}\label{e:1}
|\bag_{s'}(v_i)|=1+|\adh_{s'}(v_i)|=1+|\adh_{s'}(v_{\pi(1)})|+\wsum(x_{\pi(1)}x_{\pi(2)}\ldots x_{\pi(\pi^{-1}(i)-1)}).
\end{equation}
Observe that by construction we have
\begin{equation}\label{e:2}
\adh_{s}(v_1)=\adh_{s'}(v_{\pi(1)}).
\end{equation}
Moreover, by the definition of $\pmax(\cdot)$, Lemma~\ref{lem:reorder-two}, and Claim~\ref{cl:reshuffle-prerequisites}, we have
\begin{equation}\label{e:3}
1+\wsum(x_{\pi(1)}\ldots x_{\pi(\pi^{-1}(i)-1)})\leq \pmax(h')\leq \pmax(h)\leq \ell+1-|\adh_{s}(v_{1})|.
\end{equation}
Here, the additional $+1$ summand on the left hand side is obtained by including also the first $+$ symbol at the front of $x_i=x_{\pi(\pi^{-1}(i))}$.
By combining~\eqref{e:1} with~\eqref{e:2} and~\eqref{e:3}, we infer that $|\bag_{s'}(v_i)|\leq \ell+1$, as requested.
\cqed\end{proof}

The whole construction was set up in order to make sure that after the reorganization, the red vertices within each factor of $\Ff$
can be grouped into a small number of $U$-factors. We now check this formally.

\begin{clm}\label{cl:context-fewU}
Let $B'\in \Ff$. Then the set $B'\cap U$ can be partitioned into at most $(5\ell+3)/2$ sets that are $U$-factors in $F'$.
\end{clm}
\begin{proof}
Let us adopt the notation from the description of the reorganization of the factor $B'$.
Take any monochromatic block $x_{\pi(i)}x_{\pi(i+1)}\ldots x_{\pi(j)}$ in $h'$. 
Observe that the corresponding vertex set $C_{\pi(i)}\cup C_{\pi(i+1)}\cup \ldots \cup C_{\pi(j)}$ is a monochromatic context factor in $F'$ of the same color as the block.
Consequently, since there are at most $(5\ell+3)/2$ maximal red blocks in $h'$, the set $B'\cap U$ can be partitioned into at most $(5\ell+3)/2$ $U$-factors in $F'$.
\cqed\end{proof}

We now argue that the forest $F'$ has all the required properties.
By Claim~\ref{cl:reorganize-correct}, $F'$ is indeed an elimination forest of $G$, and 
by Claim~\ref{cl:reorganize-reduced} it is reduced.
By Claim~\ref{cl:reorganize-width}, the width of $F'$ is not larger than the width of $F$.
We now bound the number of maximal $U$-factors in $F'$. Observe that for every forest factor $A\in \refine_F(U\cup W)$, $A\cap U$ is either empty or a forest factor in $F$, which stays intact in~$F'$ (Claim~\ref{cl:forest-fewU}). On the other hand, if $B\in \refine_F(U\cup W)$ is a context factor, then $\Ff_B$ contains at most $\ell+1$ important factors and at most $\ell+1$ regular factors (Claim~\ref{cl:important-bound}). Each important factor of $\Ff_B$ is monochromatic, while for each regular factor $B'\in \Ff_B$, the set $B'\cap U$ can be partitioned into at most $(5\ell+3)/2$ $U$-factors in $F'$ (Claim~\ref{cl:context-fewU}). By Claim~\ref{cl:UW-small-number} and since $\ell\leq k$, we conclude that $U$ can be partitioned into at most
$$f(k):=(k+2)+(2k+1)\cdot (k+1)\cdot (5k+3)/2$$
$U$-factors in $F'$. Since each $U$-factor is contained in some maximal $U$-factor, and maximal $U$-factors in $F'$ form a partition of $U$ by Lemma~\ref{lem:coarsest}, we infer
that there are at most $f(k)$ maximal $U$-factors in~$F'$. This establishes condition~\ldlref{c:fewU}. 
Finally, as we said before, the satisfaction of conditions~\ldlref{c:Ucont} and~\ldlref{c:Wcont} follows easily from preserving invariants \invref{i:parent-child}--\invref{i:leaf}, 
and we leave this verification to the reader.
This concludes the proof of the Local Dealternation Lemma.

\section{Normal form for \mso transductions}\label{sec:normalization}

In this section we prove Theorem~\ref{thm:mso-normalization}.
Let us first discuss the proof strategy.
Recall that an \mso transduction is a finite sequence of {\em{atomic steps}}, each being filtering, universe restriction, interpretation, copying, or coloring.
Hence, the idea is to show that one can appropriately swap and merge these steps while modifying them slightly, so that the final normal form is achieved.
It will be trivial to implement the rules algorithmically, hence we focus only on their description.
As most of the rules are very simple, we keep the argumentation concise.

We start with merging rules: whenever two steps of the same type, apart from coloring, appear consecutively in the sequence, then they can be merged into one step.

\begin{clm}\label{cl:merge}
If $\Ii_1$ and $\Ii_2$ are two atomic transductions of the same kind, being either renaming, copying, filtering, universe restriction, or interpretation, then $\Ii_2\circ \Ii_1$ can be expressed as a 
single step of the same kind.
\end{clm}
\begin{proof}
For copying and renaming the claim is trivial.
For filtering, it suffices to take filtering with \mso sentence $\psi_1\wedge \psi_2$, where $\psi_1$ and $\psi_2$ are sentences used in $\Ii_1$ and $\Ii_2$, respectively.
For universe restriction, suppose $\varphi_1(u)$ and $\varphi_2(u)$ are two \mso formulas used in $\Ii_1$ and $\Ii_2$, respectively. Then it suffices to take a universe
restriction step using the formula $\varphi(u)=\varphi_1(u)\wedge\varphi'_2(u)$, where $\varphi'_2(\cdot)$ is constructed from $\varphi_2(\cdot)$ by restricting the 
universe to the elements satisfying $\varphi_1(\cdot)$, that is,
adding a guard to each quantifier that restricts its range to (sets of) elements satisfying $\varphi_1(\cdot)$.
Finally, for interpretation, it suffices to replace each relation atom $R(x_1,\ldots,x_r)$ appearing in each \mso formula used in $\Ii_2$, by the \mso fomula $\varphi_R(x_1,\ldots,x_r)$ used in $\Ii_1$ to
define the interpretation of $R$. The formulas obtained in this way define an interpretation that is equivalent to $\Ii_2\circ \Ii_1$.
\cqed\end{proof}

Next, we give swapping rules that enable us to exchange pairs of consecutive transductions.
We first check that renaming steps can be swapped with any other step, thus they can be always pushed to the left.

\begin{clm}\label{cl:*-rename}
Suppose $\Ii_1$ is a renaming step and $\Ii_2$ is an atomic transduction that is not renaming.
Then $\Ii_2\circ\Ii_1=\Ii'_1\circ\Ii'_2$, where $\Ii'_1$ is a renaming step and $\Ii_2'$ is an atomic step of the same kind as $\Ii_2$.
\end{clm}
\begin{proof}
If $\Ii_2$ is an interpretation step, then we can just apply Claim~\ref{cl:merge} to merge $\Ii_1$ and $\Ii_2$ into a single interpretation step $\Ii'_2$, and take $\Ii'_1$ to be identity.
For other kinds of transductions, it is trivial to rewrite $\Ii_2$ in the vocabulary before renaming, thus obtaining $\Ii'_2$, and we put $\Ii'_1=\Ii_1$.
\cqed\end{proof}

Next, we show that the universe restriction steps can be pushed to the left by swapping.

\begin{clm}\label{cl:*-restrict}
Suppose $\Ii_1$ is a universe restriction step and $\Ii_2$ is an atomic transduction that is not a universe restriction.
Then $\Ii_2\circ\Ii_1=\Jj\circ \Ii'_1\circ\Ii'_2$ for some $\Jj$, $\Ii'_1$ and $\Ii'_2$, such that $\Jj$ is a renaming step, $\Ii'_1$ is a universe restriction step, and
$\Ii'_2$ is an atomic transduction of the same kind as $\Ii_2$.
\end{clm}
\begin{proof}
Let $\varphi(\cdot)$ be the formula used by $\Ii_1$ to restrict the universe.
We proceed by case study, depending on the kind of $\Ii_2$.

If $\Ii_2$ is a coloring step, then we can take $\Ii'_1=\Ii_1$, $\Ii'_2=\Ii_2$, and $\Jj$ to be identity, since introducing the new color has no effect on the application of universe restriction.

If $\Ii_2$ is a filtering step, say using an \mso sentence $\psi$, then we can take $\Ii'_1=\Ii_1$ and $\Ii'_2$ to be filtering using $\psi$ restricted to the elements satisfying $\varphi(\cdot)$. That is,
we modify $\psi$ by adding a guard to each quantifier that restricts its range to (sets of) elements satisfying $\varphi(\cdot)$. For $\Jj$ we can take the identity.

If $\Ii_2$ is a copying step, then we can take $\Ii'_2=\Ii_2$ and $\Jj$ to be identity, and we define $\Ii'_1$ as follows.
First, let $\varphi'(u)$ be the sentence over the vocabulary after copying, obtained from $\varphi(u)$ by additionally requiring that $u$ belongs to the first layer of copies and restricting the range
of each quantifier to the first layer.
Then, $\Ii'_1$ is universe restriction with the \mso predicate $\varphi''(u)$ saying that the unique element $u'$ that is a copy of $u$ from the first layer satisfies $\varphi'(u')$.
Thus, $\varphi'(\cdot)$ works on the first layer in exactly the same manner as $\varphi(\cdot)$ worked on the original universe, 
while $\varphi''(\cdot)$ removes all copies of all elements that would be removed by $\varphi(\cdot)$.

Finally, if $\Ii_2$ is an interpretation step, then we proceed as follows. As $\Ii'_2$ we take $\Ii_2$ restricted to the elements that satisfy $\varphi(\cdot)$; that is, in every formula used in
$\Ii_2$ we restrict both the free variables and all quantifiers to elements satisfying $\varphi(\cdot)$. Moreover, $\Ii'_2$ only adds relations to the structure, interpreted via formulas modified as in the previous
sentence, while all relations of the original vocabulary are kept intact via identity interpretations. Next, we take $\Ii'_1=\Ii_1$; note that thus $\Ii'_1$ works on the original relations that were kept in
the structure. Finally, we add a renaming step $\Jj$ that removes the relations of the original vocabulary and renames the other relations added by $\Ii'_2$ to their final names.
\cqed\end{proof}

Next, we push the interpretation steps to the left by swapping.

\begin{clm}\label{cl:*-interprete}
Suppose $\Ii_1$ is an interpretation step and $\Ii_2$ is an atomic transduction, being either coloring, filtering, or copying.
Then $\Ii_2\circ\Ii_1=\Ii'_1\circ\Ii'_2$, where $\Ii'_1$ is an interpretation step and $\Ii'_2$ is an atomic transduction of the same kind as $\Ii_2$.
\end{clm}
\begin{proof}
We proceed by case study, depending on the kind of $\Ii_2$.

If $\Ii_2$ is a coloring step, then we can simply put $\Ii'_1=\Ii_1$ and $\Ii'_2$ to be $\Ii_2$ enriched by keeping the unary predicate introduced by $\Ii_1$ intact.

If $\Ii_2$ is a filtering step, say using an \mso sentence $\psi$, then we can put $\Ii'_1=\Ii_1$ and $\Ii_2$ to be a filtering using the sentence $\psi'$ obtained from $\psi$
by replacing each relation atom $R(x_1,\ldots,x_r)$ by its interpretation $\varphi_R(x_1,\ldots,x_r)$ under $\Ii_1$.

Finally, if $\Ii_2$ is a copying step, then we take $\Ii'_2=\Ii_2$ and $\Ii'_1$ defined as follows.
First, from each formula $\varphi_R(x_1,\ldots,x_r)$ used by $\Ii_1$ construct a formula $\varphi'_R(x_1,\ldots,x_r)$ by restricting all free variables and ranges of all quantifiers to the first layer
of copies. Then, in $\Ii'_2$ to interpret relation $R$ use the formula $\varphi''_R(x_1,\ldots,x_r)$ which expresses the following: the unique elements $x'_1,\ldots,x'_r$ that are the first-layer copies
of $x_1,\ldots,x_r$, respectively, satisfy $\varphi'_R(x'_1,\ldots,x'_r)$.
\cqed\end{proof}

The next type to tackle is copying.

\begin{clm}\label{cl:*-copy}
Suppose $\Ii_1$ is a copying step and $\Ii_2$ is an atomic transduction, being either filtering or coloring.
Then $\Ii_2\circ\Ii_1=\Jj\circ \Ii_1\circ\Ii'_2$, where $\Jj$ is a single interpretation step, while $\Ii'_2$ is a single filtering step if $\Ii_2$ was filtering, and $\Ii'_2$ is a finite
sequence of coloring steps if $\Ii_2$ was coloring.
\end{clm}
\begin{proof}
Let $\Ii_1$ copy the universe $\ell$ times.

First, suppose $\Ii_2$ is a filtering step, say using an \mso sentence $\psi$.
Then we can take $\Jj$ to be the identity, while $\Ii_2$ is filtering using a sentence $\psi'$ obtained from $\psi$ by restricting the ranges of all quantifiers to the first layer of copies.

Second, suppose $\Ii_2$ is a coloring step, say introducing a unary predicate $X$.
Then we take $\Ii_2'$ to be a sequence of $\ell$ coloring steps as follows. The $i$th coloring step introduces a unary predicate $X_i$. After performing the copying (transduction $\Ii_1$),
we add an additional interpretation step $\Jj$ that introduces the unary predicate $X$ interpreted as follows: if $u$ is from the $i$th layer of copies, then $u$ is declared to belong to $X$
if and only if it belongs to $X_i$; this can be easily expressed in \mso. The auxiliary predicates $X_1,\ldots,X_\ell$ are dropped by interpretation $\Jj$.
\cqed\end{proof}

Finally, we are left with swapping coloring and filtering.

\begin{clm}\label{cl:color-filter}
Suppose $\Ii_1$ is a filtering step and $\Ii_2$ is a coloring step.
Then $\Ii_2\circ\Ii_1=\Ii_1\circ\Ii_2$.
\end{clm}
\begin{proof}
The filtering may just ignore the new predicate introduced by the coloring.
\cqed\end{proof}

We now show that using the merging and swapping rules described in the above claims, we can reduce any sequence of atomic transductions to the normal form described in the theorem statement.

First, observe that by iteratively using Claim~\ref{cl:merge} (for renaming) and Claim~\ref{cl:*-rename} we can always move any renaming steps to the left of the current sequence of transductions,
and merge it there into a single renaming step. 

Next, consider the universe restriction steps. Using Claim~\ref{cl:merge} (for universe restriction) and Claim~\ref{cl:*-restrict} we can iteratively
move any universe restriction steps to the left and merge them into one universe restriction step, placed immediately to the right of the final renaming step.
Any additional renaming steps obtained during this procedure can be again pushed to the left as in the previous paragraph.

Thus, the remaining sequence has no universe restriction steps. Observe that now all interpretation steps can be moved to the left using Claim~\ref{cl:*-interprete}, and merged
into one interpretation step using Claim~\ref{cl:merge} for interpretation. This step is placed immediately to the right of the universe restriction step obtained in the previous paragraph.

We are left with a sequence consisting only of copying, filtering, and coloring steps. Apply Claim~\ref{cl:*-copy} iteratively to move all copying steps to the left, and apply Claim~\ref{cl:merge} (for copying)
to merge them into a single copying step, placed immediately to the right of the interpretation step obtained in the above paragraph. 
Any additional interpretation steps obtained during this procedure can be pushed left as above, and merged into the interpretation step obtained above.
Note that this operation may blow up coloring steps into finite sequences of coloring steps, but this is irrelevant for the pushing procedure.

Finally, we are left with filtering and coloring steps that can be sorted using Claim~\ref{cl:color-filter}. Then, filtering steps can be merged into one filtering step using Claim~\ref{cl:merge} (for filtering).
We have obtained the normal form as described in the theorem statement, hence we are done.

\section{Conclusions}\label{sec:conc}

In this work we have constructed an \mso transduction that, given a constant-width tree decomposition of a graph, computes a tree decomposition of this graph of optimum width.
As we have shown, this transduction can be conveniently composed with the \mso transduction given in~\cite{bojanczyk2016definability} to prove that given a graph of constant treewidth,
some optimum-width tree decomposition can be computed by means of an \mso transduction.

One direct application of this result is a strengthening of the main result of~\cite{bojanczyk2016definability}. 
There, we have proved that if a class of graphs of treewidth at most $k$ is recognizable (see~\cite{bojanczyk2016definability} for omitted definitions), then it can be defined in \mso with modular counting predicates.
The main technical component of this proof was Theorem 2.4, which states that for every $k$ there is an \mso transduction from graphs to tree decompositions, which given a graph of treewidth $k$ outputs
some its tree decomposition of width bounded by $f(k)$, for some doubly-exponential function $f$. 
Then the proof of the main result of~\cite{bojanczyk2016definability} used $f(k)$-recognizability, i.e., recognizability within the interface (sourced) graphs with at most
$f(k)$ interfaces (sources). By replacing the usage of Theorem 2.4 of~\cite{bojanczyk2016definability} with Corollary~\ref{cor:courcelle-stronger} of this paper, we 
deduce that only $k$-recognizability of a class of graphs of treewidth at most $k$ is sufficient to prove that it can be defined in \mso with modular counting predicates.
However, this strengthening was already known: Courcelle and Lagergren~\cite{CourcelleL96} proved that if a class of graphs of treewidth at most $k$ is $k$-recognizable, then it is
also $k'$-recognizable for all $k'\geq k$. In fact, the proof technique of Courcelle and Lagergren essentially uses the same technique as Bodlaender and Kloks~\cite{BodlaenderK96} and as we do in this work;
the main technical component of~\cite{CourcelleL96} can be interpreted as a variant of our Local Dealternation Lemma.

In Section~\ref{sec:implementing} we have seen that the algorithmic result of Bodlaender and Kloks~\cite{BodlaenderK96} can be derived from the existence of an \mso transduction solving the same task and
the fact that \mso transductions can be implemented efficiently on structures of bounded treewidth (Theorem~\ref{thm:implement-transduction}).
We believe that this approach might be applicable to other structural decompositions of graphs as well.
More precisely, suppose that we have some notion of a tree-like decomposition of a graph, where width-$k$ decompositions can be interpreted in forests labeled by an alphabet depending on $k$.
Then, the analogue of the problem of Bodlaender and Kloks would be as follows: given some (possibly suboptimal) decomposition of width $k$, compute a decomposition of optimum width.
Theorem~\ref{thm:implement-transduction} reduces this algorithmic task to showing that some optimum-width decomposition can be constructed from the input suboptimal one by means of an \mso transduction.
For this, however, we essentially only need to solve the combinatorial core: prove the analogues of Dealternation and Conflict Lemmas. The tedious algorithmic layer of designing a dynamic programming procedure is thus
abstracted away thanks to Theorem~\ref{thm:implement-transduction}, and we are left with a purely combinatorial task.

Two concrete width measures of graphs for which we believe that this algorithmic technique may be applicable are {\em{carvingwidth}} and {\em{tree-cutwidth}}.
Both of them can be seen as analogues of treewidth for edge cuts instead of vertex cuts.
For a closely related parameter {\em{cutwidth}}, Giannopoulou et al.~\cite{GiannopoulouPRT16} have very recently given a linear-time fixed-parameter algorithm based on
combining an analogue of the Bodlaender-Kloks algorithm~\cite{BodlaenderK96} with an analogue of the reduction scheme of Bodlaender~\cite{Bodlaender96}.
It is conceivable that a similar approach can be applied to carvingwidth and tree-cutwidth, where the analogue of Bodlaender-Kloks algorithm would be obtained via implementing the task by an \mso transduction,
whereas for the reduction scheme one would use a similar scheme as in~\cite{GiannopoulouPRT16}.
This would yield linear-time fixed-parameter algorithms for computing optimum-width carving and tree-cut decompositions.
To the best of our knowledge, 
no such algorithm are known yet: Giannopoulou et al.~\cite{GiannopoulouKRT19} gave a linear-time fixed-parameter algorithm to determine the optimum value of the tree-cutwidth, which however does not provide a witnessing decomposition, and we are not aware of analogous algorithmic developments for carvingwidth.

\paragraph*{Acknowledgements.} The authors would like to thank Bruno Courcelle for pointing out connections with his work with Jens Lagergren~\cite{CourcelleL96}.

\bibliographystyle{alphaurl}
\bibliography{kloks}

\end{document}